%% file: main.tex
\documentclass[11pt]{article}

\input{preamble}

\title{Estimating Discrete Games of Complete Information: \\ Bringing Logit Back in the Game\footnote{I thank Matthew Backus, Arie Beresteanu, Gautam Gowrisankaran, Gaston Illanes, Karam Kang, Sokbae Lee, Qingmin Liu, Bernard Salanié, Philipp Schmidt-Dengler, and the seminar participants at Columbia University, U.S. Federal Trade Commission, Yonsei University, the 2025 World Congress of the Econometric Society, and Korean Econometric Society for helpful discussion and comments. This paper is based on the third chapter of my Ph.D. dissertation. A portion of this paper was completed during the author’s tenure at the U.S. Federal Trade Commission; the views expressed here are solely those of the author and do not necessarily reflect those of the Commission or its Commissioners. All errors are mine.}}

\author{Paul S. Koh\footnote{School of Economics, Yonsei University. Email: \texttt{pskoh@yonsei.ac.kr}.}}
\date{June 10, 2026}

\begin{document}

\maketitle
\input{abstract}

\input{S.01.Introduction}
\input{S.02.Model}

\input{S.03.Convex.Outer.Set}
\input{S.04.Unordered.Actions}
\input{S.05.OrderedActions}
\input{S.07.Empirical.Application}

\input{S.08.Empirical.Application.II}

\input{S.99.Conclusion}

%%% REFERENCE %%%
\singlespacing
% \clearpage
\bibliographystyle{econ}
\bibliography{references}
\doublespacing

% \clearpage
\part*{Appendix}
\appendix
\input{S.A.Proofs}

\clearpage
\begin{center}
{\Large\bfseries Online Appendix for \\
``Estimating Discrete Games of Complete Information:\\
Bringing Logit Back in the Game''
}
\end{center}
\pagenumbering{arabic}
\setcounter{page}{1}
\input{S.B.Extension.to.Games.with.Vector.Decisions}
\input{S.C.Additional.Illustrations.for.Ordered.Actions}

\input{S.D.Computational.Details}

% \clearpage
% \part*{Estimation and Inference in Games with Ordered Actions}
% \input{S.X.Tractable.Sharp.Identified.Set.Under.Ordered.Actions}
% \input{S.Y.Revoming.Redundant.Inequalities}
% \clearpage
% \input{S.Z.Misc}
% \input{S.Z.Referee.Comments}

\end{document}

%% file: preamble.tex
\usepackage{graphicx} % Required for inserting images
\usepackage[utf8]{inputenc}
\usepackage{eurosym,geometry,graphicx,color,setspace,sectsty,comment,footmisc,natbib,pdflscape,array,hyperref}
\usepackage[normalem]{ulem}
\usepackage{caption}
\usepackage{subcaption}
\usepackage{amssymb,amsfonts,amsmath}
\usepackage{amsthm}
\usepackage{natbib}
\usepackage{booktabs}
\usepackage{threeparttable}
\usepackage{makecell}
\usepackage{tikzsymbols} % emoji
\usepackage{tikz}
\usetikzlibrary{patterns,arrows,positioning,fit,shapes.geometric,backgrounds}
\usepackage[capposition=top]{floatrow} % For Figure notes

\usepackage{xcolor} % Colored text
\usepackage{todonotes}
\setuptodonotes{inline}
\usepackage{mdframed}
\newtheorem{assumption}{Assumption}
\newtheorem{lemma}{Lemma}
\newtheorem*{lemma*}{Lemma}
\newtheorem{theorem}{Theorem}
\newtheorem*{theorem*}{Theorem}

\theoremstyle{definition} % Upright text

\newtheorem{example}{Example}
\newtheorem*{example*}{Example}

\AtBeginEnvironment{example}{%
  \pushQED{\qed}%
}
\AtEndEnvironment{example}{\popQED\endexample}

\AtBeginEnvironment{remark}{%
  \pushQED{\qed}%
}
\AtEndEnvironment{remark}{\popQED\endexample}

\hypersetup{colorlinks,citecolor=blue} % citation in blue

%% file: abstract.tex
\begin{abstract}
Discrete games are central tools for empirical analysis of strategic interaction, but equilibrium multiplicity and partial identification often make them computationally difficult to estimate. This paper develops tractable methods for estimation and inference in complete-information discrete games. The key idea is to construct an outer set by comparing observed frequencies of action profiles with singleton-class generalized likelihoods: model-implied probabilities that those profiles can arise as equilibria. The resulting conditional moment inequalities avoid computationally expensive equilibrium enumeration, numerical simulation, and grid search. Under standard empirical assumptions used in discrete-game models, including logit payoff shocks, these restrictions have closed-form expressions and are convex in a subvector of structural parameters. I develop the approach for both unordered and ordered action spaces. Monte Carlo experiments and empirical applications show that the methods deliver informative outer sets and can reduce computation time by several orders of magnitude relative to existing approaches.

\phantom{skip} \\
\textbf{Keywords}: Discrete games, complete information, partial identification, logit, convex program %\\
%\textbf{JEL Codes:} C13, C57, L10
\end{abstract} 
\clearpage

%% file: S.01.Introduction.tex
\section{Introduction}

Econometric models of strategic interaction have become standard tools in empirical industrial organization and related fields \citep{tamer_incomplete_2003, de2013econometric, aradillas-lopez_econometrics_2020, kline2021moment}. Yet computation remains a central obstacle, especially in complete-information discrete games. These models are typically partially identified because the same primitives may support multiple equilibria and the econometrician usually does not observe the equilibrium-selection rule. One approach is to restore point identification by imposing additional structure on equilibrium selection, but such structure is often difficult to discipline empirically. Partial-identification methods avoid specifying an equilibrium-selection rule, but their implementation can be computationally demanding, often requiring repeated equilibrium computation, simulation, or high-dimensional grid search. \citet{ciliberto2021superstar}, for example, report that their implementation took about a week. More broadly, leading researchers in the field view computational burden as a central obstacle to the wider adoption of moment-inequality methods in empirical work \citep{kline2021moment, canay2023user}.

This paper develops a simple and scalable approach to estimating a large class of finite, static, discrete games of complete information. The framework accommodates unordered and ordered actions, market-level unobserved heterogeneity, pure-strategy Nash equilibrium, and unrestricted equilibrium selection. The main idea is to construct generalized-likelihood-based conditional moment inequalities that, under standard log-concavity conditions on players' payoff shocks, are convex in a subvector of payoff parameters. This convexity allows the researcher to compute informative outer sets by solving convex projection problems over payoff parameters, rather than by extensive equilibrium enumeration, simulation over idiosyncratic payoff shocks, or high-dimensional grid search. Logit-type assumptions on payoff shocks further enhance tractability by yielding closed-form expressions. Thus, the analytic convenience familiar from single-agent discrete choice can persist in multi-agent complete-information settings when the researcher uses appropriately chosen generalized likelihoods.

The starting point is the sharp characterization of \citet{galichon2011setidentification}, which I review in Section~\ref{section:econometric.problem}. In their framework, the identified set is characterized by conditional moment inequalities requiring the observed conditional probability of each event to be no larger than its generalized likelihood: the largest probability that the model can assign to the event over all admissible equilibrium-selection rules. This characterization delivers sharp identification without specifying an equilibrium-selection rule, but its implementation in complete-information discrete games remains computationally burdensome because generalized likelihoods must be evaluated repeatedly across events, covariate values, and parameter values.

In Section \ref{section:convex.outer.set}, I propose selecting a \emph{subset} of these identifying inequalities and define an \emph{outer set} characterized by conditional moment inequalities of the form
\begin{equation}\label{equation:intro.key.idea}
    m_j(\theta) \leq 0, \quad j \in \mathcal{J},
\end{equation}
where $\theta$ is the structural parameter vector. The key contribution is to show how to choose $\mathcal{J}$ so that each $m_j$ is convex in a payoff-parameter subvector $\gamma$ of $\theta=(\gamma,\sigma)$, where $\sigma$ governs the distribution of common market-level unobservables. Without common unobservables, the resulting outer set is convex in the full parameter vector $\theta$. With common unobservables, Pr\'ekopa's theorem implies convexity in $\gamma$ conditional on the common-shock scale parameter $\sigma$, so the outer set can be computed as a union of convex slices indexed by $\sigma$.

The construction exploits the best-response structure of Nash equilibrium. For a candidate action profile \(y=(y_1,\ldots,y_I)\), the event that \(y\) is a pure-strategy Nash equilibrium is the event that each player \(i\) finds \(y_i\) optimal when opponents' actions are fixed at \(y_{-i}\). Combining this observation with the generalized-likelihood approach decomposes the singleton-event generalized likelihood, conditional on the common shock, into a product of single-agent best-response probabilities. Under affine payoff differences and log-concave unobservables, these best-response probabilities are log-concave; multiplying across players and integrating over the log-concave common shock preserve log-concavity in the payoff parameters. This is the source of the convexity and computational tractability of the proposed outer set.

Logit assumptions further strengthen this tractability by giving closed-form expressions for the player-level best-response probabilities. For unordered action spaces, these terms reduce to multinomial-logit probabilities (Section~\ref{section:unordered.actions}). For ordered action spaces, the ordered-response threshold structure yields ordered-logit interval probabilities (Section~\ref{section:ordered.actions}). Thus, the method uses the logic of Nash best responses to turn a multiple-equilibrium complete-information problem into a collection of single-agent probability calculations, restoring much of the computational power of logit without imposing an equilibrium-selection rule.

I also provide a computationally simple approach to inference. I account for sampling uncertainty in the first-step estimates of conditional choice probabilities by constructing simultaneous one-sided lower confidence bounds. These lower bounds can be substituted directly into the moment inequalities, yielding confidence sets that remain conservative while preserving the convex-programming structure used to estimate the identified set.

I illustrate the framework with two empirical applications. In Section \ref{section:empirical.application}, I first revisit the binary entry game between Walmart and Kmart, following \citet{ellickson_structural_2011}'s simplified version of \citet{jia_what_2008}'s model. I use this application because it is a canonical unordered-action entry game that can be implemented with publicly available data, making it a transparent benchmark for the proposed procedure. The singleton-event outer-set procedure computes informative projection intervals for nine parameters in \(112.3\) seconds of total computational time. As a conservative comparison, I then compute a sampled approximation to the sharp confidence set only after using this first-step outer set to narrow the search region. Even on this reduced region, the adaptive sharp-set search takes \(3.9\) worker-hours, or \(14{,}060.9\) seconds, of total distributed computational time to collect \(10\) million accepted sharp/core points in my implementation. These accepted-point ranges are still numerical approximations rather than exact sharp-set projections; a more exhaustive search for boundary points would further increase the computational cost. Thus, the comparison is conservative in favor of the sharp-set search and likely understates the cost of exact sharp-set computation. Even under this favorable benchmark, the convex singleton-event outer-set procedure is more than two orders of magnitude faster while delivering informative bounds.

The second application in Section~\ref{section:empirical.application.II} studies entry by McDonald's, Burger King, and Wendy's in an ordered-action game where each chain can operate zero, one, or two or more outlets in a market. I use this application to illustrate how the approach scales to games with many possible outcomes and, consequently, many potential equilibria---a setting in which direct sharp-set computation or point estimation based on explicit equilibrium-selection assumptions would be substantially more difficult. The ordered-action procedure computes projection intervals for \(18\) normalized payoff parameters in \(7.8\) worker-hours of total distributed computational time, corresponding to \(78.9\) minutes of elapsed wall-clock time on six Julia worker processes. The estimates are economically intuitive: local restaurant activity raises payoffs, payoffs are concave in own outlet counts, and rival outlets have adverse effects on profitability. The own-network effects are less precisely identified, while the accepted common-shock values imply substantial correlation in chains' unobserved market-level payoff shocks.

\subsection{Related Literature}

This paper contributes to the literature on complete-information discrete games
with multiple equilibria, including \citet*{tamer_incomplete_2003},
\citet*{andrews_confidence_2004}, \citet*{ciliberto2009marketstructure},
\citet*{bajari2010identification}, \citet*{galichon2011setidentification},
\citet*{beresteanu_sharp_2011}, \citet*{henry2015combinatorial},
\citet*{pakes_moment_2015}, \citet*{kline_bayesian_2016},
\citet*{aradillas2022inference}, and \citet*{koh2023stable}. This literature
provides sharp characterizations and inference procedures for incomplete models
generated by multiple equilibria. Direct implementation, however, can be
computationally demanding because it often combines equilibrium computation,
probability approximation over payoff shocks and equilibrium regions, and
parameter-space search. My contribution is to reformulate a useful subset of
generalized-likelihood inequalities so that they generate convex restrictions
in payoff parameters. I focus on singleton equilibrium events: the observed
probability of an action profile cannot exceed the probability that the profile
is an equilibrium. These restrictions are generally non-sharp, but they retain
direct equilibrium content. Under affine payoff differences and log-concave
shocks, the corresponding generalized likelihoods are log-concave; under
logit-type shocks, the player-level best-response probabilities have
closed-form expressions. The resulting outer set can therefore be computed
through convex projection problems over payoff parameters, rather than through
high-dimensional grid search over the full parameter space.

The closest recent contribution is \citet{fan2025estimating}, who develop a
scalable outer-set approach for games with many firms and many discrete
decisions, with a leading formulation in which firms choose vectors of binary
actions. Their key insight is to use dominance and non-dominance bounds on
firms' conditional choice probabilities, which are easy to evaluate and
especially powerful in product-portfolio settings with many binary entry
decisions. My approach is complementary. Rather than bounding marginal binary
choice probabilities, I use singleton-event generalized likelihoods: the
observed probability of an action profile cannot exceed the probability that
the profile is a pure-strategy Nash equilibrium. This event-level construction
applies directly to finite unordered action spaces and ordered action spaces.
It also delivers a different computational advantage: under affine payoff
differences and log-concave unobservables, the resulting generalized likelihoods
are log-concave in payoff parameters, so projection intervals can be computed
by convex optimization rather than by stochastic search over candidate
parameter values. Market-level common unobserved heterogeneity is built into
the framework: conditional on the common shock, singleton generalized
likelihoods factor into products of single-agent best-response probabilities,
and integrating over the common shock preserves log-concavity. Because the two
approaches use different event classes and exploit different forms of structure,
neither set of inequalities generally nests the other. This paper is also
related to \citet{gowrisankaran2024computable}, who use monotonicity in private
cost shocks to compute dynamic oligopoly models with many ordered
capacity-investment choices. Their approach and mine share the goal of using
economic structure to avoid brute-force equilibrium computation in large
discrete games, but the objects differ: they study computation of dynamic
oligopoly models with ordered investment choices, whereas I construct convex
generalized-likelihood inequalities for static complete-information games with
unrestricted equilibrium selection.

The paper also contributes to the literature on moment inequalities, partial
identification, and tractable outer sets. Existing work develops inference for
conditional moment inequalities
\citep*{andrews2013inference,chernozhukov2013intersection,lee2013testing,
armstrong2015asymptotically,lee2018testing} and shows how linearity or
convexity can simplify estimation and inference
\citep*{beresteanu2008asymptotic,kaido2014asymptotically,gafarov2019simple,
andrews2023inference,cho2024simple}. In applied work, moment inequalities are
often chosen not only for sharpness but also for empirical content and
computational feasibility; the revealed-preference inequalities of
\citet{pakes_alternative_2010} and \citet*{pakes_moment_2015} are leading
examples. The outer sets proposed here should be understood in this
computational tradition. They deliberately omit some valid restrictions to
preserve convexity, but they retain direct equilibrium content, can be
informative on their own, and can serve as first-stage screens for sharper but
more expensive procedures. This interpretation is consistent with the caution
emphasized by \citet{li2024discordant}: non-sharp implications should be chosen
transparently, labeled as non-sharp, and interpreted with care, especially under
misspecification. In the present setting, sharp characterizations remain
available through existing methods, including \citet{beresteanu_sharp_2011},
\citet{galichon2011setidentification}, \citet*{henry2015combinatorial},
\citet{chesher2020econometric}, \citet{koh2023stable}, and
\citet*{luo2024selecting}. The contribution is not to replace sharp
identification conceptually, but to make empirically useful outer sets and
subsequent sharp-set searches computationally feasible in settings where direct
sharp-set implementation is otherwise difficult.

Finally, the empirical applications relate to work on strategic entry,
location, and network expansion by retail chains. The Walmart--Kmart
application builds on the canonical retail-entry settings studied by
\citet{jia_what_2008} and \citet{ellickson_structural_2011}. The fast-food
application relates to studies of restaurant and retail-chain entry,
cannibalization, and network economies, including
\citet{toivanen2005market}, \citet{yang2012burger}, \citet{gayle2015choosing},
\citet{igami2016unobserved}, \citet{aguirregabiria2020identification},
\citet{yang2020learning}, \citet{holmes2011diffusion},
\citet*{ellickson2013estimating}, \citet{nishida2015estimating}, and
\citet{aradillas2022inference}. The ordered-action application is especially
related to \citet{aradillas2022inference}, who develop inference for
complete-information ordered-response games using shape restrictions and
adjacent-action comparisons. Their framework applies more generally to ordered action spaces, but their specialized parametric results and empirical application focus on a two-player strategic-substitutes game. My approach uses a different computational object:
singleton-event generalized likelihoods. Under logit payoff shocks, these
probabilities have ordered-logit closed forms and generate convex outer-set
restrictions. The fast-food application illustrates how this approach can be
implemented in a three-player ordered-action game with directed pairwise
competitive effects, without imposing an equilibrium-selection rule.

% \subsection{Outline}

% The rest of the paper is organized as follows. Section \ref{section:econometric.problem} introduces the econometric problem and generalized likelihoods. Section \ref{section:convex.outer.set} defines a computationally tractable outer set. Sections \ref{section:unordered.actions} and \ref{section:ordered.actions} develop the unordered- and ordered-action procedures. Sections \ref{section:empirical.application} and \ref{section:empirical.application.II} present the Walmart--Kmart and fast-food applications. Section \ref{section:conclusion} concludes. All proofs are in Appendix \ref{section:proofs}.

%% file: S.02.Model.tex
\section{Model and Setup \label{section:econometric.problem}}

This section introduces the class of finite complete-information discrete games studied in the paper and reviews the generalized-likelihood characterization of the sharp identified set proposed by \citet{galichon2011setidentification}. The key point is that the game induces an incomplete econometric model in which, at a given state, the model may predict multiple equilibrium outcomes without specifying which one is selected.

\subsection{Discrete Game and Data}

I consider static discrete games of complete information with finite players and actions. The game primitives are $\langle \mathcal{I},(\mathcal{Y}_i)_{i\in\mathcal{I}},(u_i^\theta)_{i\in\mathcal{I}}\rangle$, where  $\mathcal{I}=\{1,\ldots,I\}$ is the set of players, $\mathcal{Y}_i$ is player $i$'s finite action set, and  $u_i^\theta:\mathcal{Y}\times\Omega\to\mathbb{R}$ is player $i$'s payoff function. I write  $\mathcal{Y}\equiv \times_{i=1}^I\mathcal{Y}_i$ and represent an action profile as  $y=(y_i,y_{-i})$, where $y_{-i}$ denotes the actions of player $i$'s opponents. Players observe the realized state $\omega$, and the game is common knowledge. Parameter $\theta$ governs payoffs and the distribution of states. The solution concept is pure-strategy Nash equilibrium.\footnote{My econometric strategy does not directly apply to cases involving mixed-strategy Nash equilibria, but it can still help narrow the parameter search region if the analyst wishes to consider it.} For each state of nature \(\omega\), define the pure-strategy equilibrium correspondence
\[
G_\theta(\omega) \equiv \left\{ y\in\mathcal Y: u_i^\theta(y_i,y_{-i},\omega) \geq u_i^\theta(\tilde y_i,y_{-i},\omega) \quad \text{for all } \tilde y_i\in\mathcal Y_i \text{ and all } i\in\mathcal I \right\}.
\]
Thus, \(G_\theta(\omega)\) is the set of pure-strategy Nash equilibrium action profiles at state \(\omega\).

The model is incomplete \`a la \citet{tamer_incomplete_2003}: $G_\theta(\omega)$ may contain multiple equilibrium outcomes, and the model is silent on which one is selected. An equilibrium selection rule $\pi_\theta:\Omega\to\Delta(\mathcal{Y})$ assigns probabilities to outcomes subject to the support restriction that $\pi_\theta(y\mid\omega)>0$ only if $y\in G_\theta(\omega)$.

Let $\omega=(x,\xi)$, where $x\in\mathcal{X}$ and $\xi\in\Xi$ denote variables that are observable and unobservable to the econometrician, respectively. I assume $\mathcal{X}$ is discrete and has full support.\footnote{Assuming $\mathcal{X}$ is discrete with full support simplifies the exposition by avoiding measure-theoretic details that are typically irrelevant in empirical applications. In practice, continuous covariates are often discretized for computational tractability; see, e.g., \citet{ciliberto2009marketstructure}.} The data are generated by a true parameter $\theta_0\in\Theta$ and an equilibrium-selection rule, neither of which is known to the econometrician. The econometrician observes a cross-sectional sample $\{y_m,x_m\}_{m=1}^n$, where $m$ indexes independent observational units such as markets. Each realized outcome $y_m$ is selected from $G_{\theta_0}(\omega_m)$ according to the equilibrium selection rule.

I assume that, as $n\to\infty$, the econometrician can identify the vector of conditional choice probabilities (CCPs) $\phi:\mathcal{X}\to\Delta(\mathcal{Y})$, where each $\phi(y\mid x)\in[0,1]$ denotes the probability of observing outcome $y$ at observable state $x$. Thus, I treat $\phi$ as a known constant vector while establishing identification arguments.

\begin{example}[Two-player entry game] \label{example:two.player.entry.game} I use a two-player entry game as a running example. There are two players $i=1,2$. Each player $i$ can choose to enter ($y_i=1$) or stay out ($y_i=0$), so $\mathcal{Y}_i=\{0,1\}$ and $\mathcal{Y}=\{(0,0),(0,1),(1,0),(1,1)\}.$ Payoffs are
\begin{equation}\label{equation:entry.game.utility.specification}
u_i^\theta(y_i,y_{-i},x,\lambda,\varepsilon_i)
=
\begin{cases}
\beta_i^\top x_i+\Delta_i y_j+\sigma\lambda+\varepsilon_i(1) 
& \text{if } y_i=1,\\
\varepsilon_i(0) 
& \text{if } y_i=0.
\end{cases}
\end{equation}
Parameter $\Delta_i\in\mathbb R$ captures the competitive effect of the opponent's presence on player $i$'s profit. Random variable $\lambda\in\mathbb R$ is a market-level payoff shock common to both players, and $\varepsilon_i=(\varepsilon_i(0),\varepsilon_i(1))\in\mathbb R^2$ captures player-action-specific idiosyncratic payoff shocks; $\lambda$, $\varepsilon_1$, and $\varepsilon_2$ are unobserved by the econometrician.\footnote{For example, consider McDonald's and Burger King's entries to local markets \citep{koh2023stable}. The common shock $\lambda$ would include local consumers' general taste for fast-food burgers, whereas $\varepsilon_i$ would reflect brand-specific consumer loyalty.} Parameter $\sigma\in\mathbb R_+$ controls the importance of the common shock. I assume $\lambda$ is an independent draw from the standard normal distribution and each $\varepsilon_i(y_i)$ independently follows the Type-I extreme value distribution. The parameter vector is $\theta=(\beta_1,\beta_2,\Delta_1,\Delta_2,\sigma)$. Equivalently, player $i$ enters if and only if $x_i^\top\beta_i+\Delta_i y_j+\sigma\lambda+\epsilon_i\geq 0,$ where $\epsilon_i\equiv\varepsilon_i(1)-\varepsilon_i(0)$ follows the standard logistic distribution.

\begin{figure}[!htbp]
\centering
\includegraphics[width=0.45\linewidth]{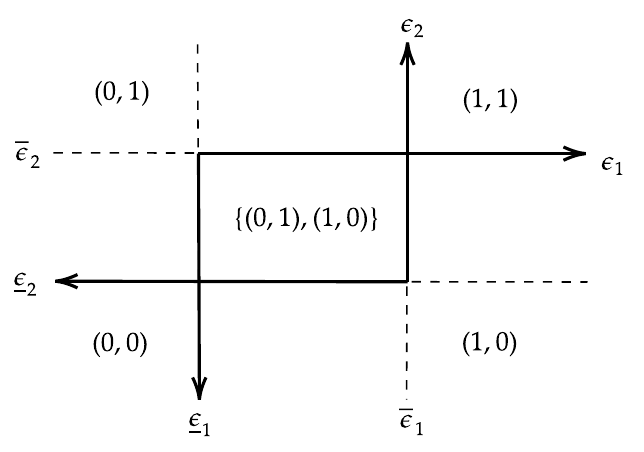}
\caption{Structure of equilibria in a two-player entry game when $\Delta_i\leq0$\label{PSNE fig:structure}}
\floatfoot{\emph{Notes}: This figure illustrates the structure of equilibria in a canonical two-player entry game when players' actions are strategic substitutes. The model predicts two equilibria ($y=(0,1)$ or $y=(1,0)$) if the players' payoff shocks $(\epsilon_1,\epsilon_2)$ are realized in the center region.}
\end{figure}

Figure \ref{PSNE fig:structure} shows the set of equilibria at each realization of $(\epsilon_1,\epsilon_2)$ when $\Delta_i\leq 0$ for all $i$. The center region admits two Nash equilibrium outcomes, $(0,1)$ and $(1,0)$; the other regions admit a unique equilibrium. The econometrician's task is to estimate $\theta$ while treating the equilibrium selection rule in the multiple-equilibrium region as unknown.
\end{example}

\subsection{Generalized Likelihood and Sharp Identification}

The sharp identified set is the collection of parameters at which the model is compatible with the observed CCPs. Since the equilibrium selection rule is unknown, the econometrician cannot reject $\theta$ if some admissible selection mechanism can generate $\phi$.

A conditional choice probability vector $q:\mathcal X\to \Delta(\mathcal Y)$ is \emph{feasible at $\theta$} if there exists an equilibrium selection rule $\pi_\theta$ such that 
\[
q(y\mid x)=\mathbb E_\theta[\pi_\theta(y\mid\omega)\mid x]
\]
for all $y\in\mathcal Y$ and $x\in\mathcal X$. Define the \emph{generalized likelihood function} as
\begin{equation}\label{equation:generalized.likelihood.function}
    \mathcal{L}_\theta(A \mid x)
    \equiv 
    \mathbb{E}\left[
    \mathbb{I}\{G_\theta(\omega)\cap A\neq\emptyset\}
    \mid x
    \right].
\end{equation}
The value $\mathcal{L}_\theta(A\mid x)$ is the largest probability that the model can assign to event $A$ conditional on $x$, because the selection rule can choose an outcome in $A$ whenever at least one such outcome is feasible. The following theorem restates Theorem 1 of \citet{galichon2011setidentification} in the present notation.

\begin{theorem}[\citet{galichon2011setidentification}] \label{theorem:GH.Theorem.1}
An arbitrary vector of conditional choice probabilities $q:\mathcal{X}\to\Delta(\mathcal{Y})$ is feasible at $\theta$ if and only if
\begin{equation} \label{equation:sharp.set.of.predictions.with.finite.inequalities}
    q(A \mid x)
    \leq
    \mathcal{L}_\theta(A \mid x),
    \quad
    \forall A\in 2^{\mathcal{Y}},\ x\in\mathcal{X},
\end{equation}
where $q(A\mid x)\equiv \sum_{y\in A}q(y\mid x)$.
\end{theorem}

Theorem \ref{theorem:GH.Theorem.1} says that a conditional choice probability vector is feasible at $\theta$ if and only if it does not contradict the maximal probabilities admissible by the model. The necessity of \eqref{equation:sharp.set.of.predictions.with.finite.inequalities} follows directly from the support restriction on the selection rule; the substantive content of \citet{galichon2011setidentification} is the converse.\footnote{To see necessity, note that a selection rule $\pi_\theta(\cdot\mid\omega)$ with support contained in $G_\theta(\omega)$ satisfies $\sum_{y\in A}\pi_\theta(y\mid\omega)\leq \mathbb I\{G_\theta(\omega)\cap A\neq\emptyset\}$ for all $A\subseteq\mathcal Y$. Taking conditional expectations yields \eqref{equation:sharp.set.of.predictions.with.finite.inequalities}.} Thus, a candidate parameter $\theta$ enters the sharp identified set if and only if the observed CCP vector is feasible at $\theta$:
\begin{equation}\label{equation:characterization.of.sharp.identified.set}
\Theta_I^\textrm{sharp} = \left\{ \theta \in \Theta: \; \phi(A \mid x)
    \leq
    \mathcal{L}_\theta(A \mid x),
    \quad
    \forall A\in 2^\mathcal{Y},\ x\in\mathcal{X} \right\}.
\end{equation}

This characterization is powerful because it eliminates the need to deal directly with the equilibrium selection rule, which may be infinite-dimensional. However, direct implementation of the sharp characterization can remain computationally difficult in parametric discrete games, because the generalized likelihood must be evaluated repeatedly across states, shocks, and candidate parameter values.

Theorem \ref{theorem:GH.Theorem.1} characterizes the sharp identified set through inequalities indexed by all subsets of $\mathcal{Y}$. In practice, the full collection may not be needed: some inequalities may be redundant, so a smaller event class can still characterize the same set. Such a lower-cardinality class is called a \emph{core-determining class}, because it suffices to characterize the core---the set of probability measures consistent with the model's set-valued predictions. This idea is studied by \citet{galichon2011setidentification} and \citet*{luo2024selecting}. My approach builds on this idea by selectively dropping inequalities to balance computational tractability with identifying power.

%% file: S.03.Convex.Outer.Set.tex
\section{Convex Outer Sets}
\label{section:convex.outer.set}

\subsection{Singleton-Event Class Outer Set}

Starting from the generalized-likelihood characterization in
\eqref{equation:characterization.of.sharp.identified.set}, I focus on the singleton-event class $\mathcal A^{*} \equiv \bigl\{\{y\}:y\in\mathcal Y\bigr\} \subseteq 2^\mathcal{Y}$. The resulting outer set is
\begin{equation}
\label{equation:singleton.event.outer.class}
\Theta_I^* = \left\{ \theta\in\Theta: \phi(y\mid x) \leq \mathcal L_\theta(y\mid x), \quad \forall y\in\mathcal Y,\ x\in\mathcal X \right\},
\end{equation}
where $\mathcal L_\theta(y\mid x)$ denotes $\mathcal L_\theta(\{y\}\mid x)$.\footnote{These restrictions are in the spirit of the singleton equilibrium-event inequalities used by \citet*{andrews_confidence_2004}. The contribution here is to show that this event class has a useful convexity structure: for a broad class of unordered- and ordered-action games, the singleton-event generalized likelihoods are log-concave and can be further simplified under standard logit assumptions.}
Because the singleton-event inequalities are a subset of the sharp generalized-likelihood inequalities, $\Theta_I^{\mathrm{sharp}} \subseteq \Theta_I^*.$ The set $\Theta_I^*$ need not be sharp, but it can be highly informative in practice and can also serve as a computational screen for sharper procedures. While the tightness of an identified set is ultimately an empirical question, I show in the simulations and applications below that the proposed outer set is often tight.

I allow for common unobserved heterogeneity.\footnote{Common unobserved heterogeneity refers to market-level payoff components observed by players but unobserved by the econometrician, such as latent demand or cost conditions. Because these components enter multiple players' payoffs, integrating them out induces correlation across players' composite payoff shocks.} Write $\theta=(\gamma,\sigma)$, where $\gamma$ collects payoff parameters and $\sigma$ collects low-dimensional parameters governing the loading of common unobserved heterogeneity. For each fixed $\sigma\in\Sigma$, define the slice
\begin{equation}\label{equation:gamma.slice}
\Gamma_I^*(\sigma) \equiv \left\{ \gamma\in\Gamma: \phi(y\mid x) \leq \mathcal L_{(\gamma,\sigma)}(y\mid x), \quad \forall y\in\mathcal Y,\ x\in\mathcal X \right\},    
\end{equation}
and let $\Theta_I^*(\sigma) \equiv \Gamma_I^*(\sigma)\times\{\sigma\}.$ Then
\begin{equation}\label{equation:outer.set.is.union.of.convex.sets}
\Theta_I^* = \bigcup_{\sigma\in\Sigma} \Theta_I^*(\sigma).    
\end{equation}
The key result below gives primitive conditions under which $\mathcal L_{(\gamma,\sigma)}(y\mid x)$ is log-concave in $\gamma$. Under those conditions, each slice $\Gamma_I^*(\sigma)$ is convex, so the outer set can be computed as a union of convex problems indexed by the low-dimensional parameter $\sigma$.

\begin{figure}[htbp!]
\centering
\caption{A non-convex outer set as a union of convex slices}
\label{figure:singleton.outer.set.convex.slices}
\includegraphics[width=0.35\textwidth]{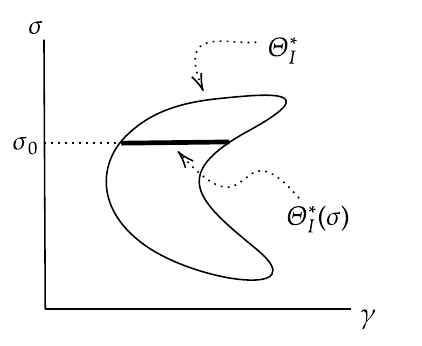}
\floatfoot{\emph{Notes}: The figure illustrates the geometry behind the
singleton-event outer set. Although the full outer set $\Theta_I^{*}$ need
not be convex, fixing the common-shock parameter $\sigma$ yields a convex
slice $\Theta_I^*(\sigma)\equiv\Gamma_I^{*}(\sigma)\times\{\sigma\}$. The full
set can therefore be computed as a union of convex problems indexed by
$\sigma$.}
\end{figure}

Figure \ref{figure:singleton.outer.set.convex.slices} illustrates this geometry. The full set $\Theta_I^*$ need not be convex because $\sigma$ is allowed to vary. However, for each fixed $\sigma$, the feasible set in $\gamma$ is convex. When common unobserved heterogeneity is absent, $\sigma$ is not needed, and the singleton-event outer set itself is convex.

The remainder of the section provides the primitive log-concavity result for generic discrete-game settings. Then, in the following sections, I specialize the player-level best-response probabilities to unordered- and ordered-action games, showing how logit assumptions further simplify computation.

\subsection{Primitive Conditions for Convex Singleton-Event Outer Sets}
\label{subsection:primitive.foundation.for.convex.singleton.event.outer.sets}

Recall that $\omega=(x,\xi)$ and write $\xi=(\lambda,\varepsilon)$, where
$\lambda\in\Lambda$ is a payoff shock common to all players and
$\varepsilon=(\varepsilon_1,\ldots,\varepsilon_I)$ collects player-specific
idiosyncratic payoff shocks. Let $G_\theta(x,\lambda,\varepsilon)$ denote the
set of pure-strategy Nash equilibria.

\begin{assumption}[Pure-strategy equilibrium]
\label{assumption:psne}
The action set $\mathcal Y_i$ is finite for each player $i$, and the solution
concept is pure-strategy Nash equilibrium.
\end{assumption}

\begin{assumption}[Independent log-concave unobservables]
\label{assumption:independent.log.concave.unobservables}
The common shock $\lambda$ and the idiosyncratic shocks
$\varepsilon_1,\ldots,\varepsilon_I$ are mutually independent and independent
of $x$. The distribution of $\lambda$ has a log-concave density $h$. For each
player $i$, $\varepsilon_i$ has a log-concave density $q_i$, and the joint
density of $\varepsilon=(\varepsilon_1,\ldots,\varepsilon_I)$ is
$q(\varepsilon)=\prod_{i=1}^I q_i(\varepsilon_i)$. None of these densities
depends on $\theta$.
\end{assumption}

\begin{assumption}[Additive unobservables and linear payoffs]
\label{assumption:additive.separability.linear.payoffs}
The parameter vector can be partitioned as $\theta=(\gamma,\sigma)$, where $\gamma$ collects payoff parameters and $\sigma$ collects parameters governing the loading of common unobserved heterogeneity. For each fixed $\sigma\in\Sigma$, each player's payoff can be written as
\[
u_i^{(\gamma,\sigma)}(y,x,\lambda,\varepsilon_i) = v_i^\gamma(y,x) + r_i(y_i)^\top S_i(\sigma)\lambda + r_i(y_i)^\top \varepsilon_i,
\]
where $r_i(y_i)$ is known, $S_i(\sigma)$ is fixed when $\sigma$ is fixed, and $v_i^\gamma(y,x)$ is affine in $\gamma$ for each $i$, $y$, and $x$.
\end{assumption}

Assumption \ref{assumption:psne} restricts attention to pure-strategy Nash equilibria. This restriction is substantive, because the econometric strategy does not directly extend to mixed-strategy equilibria.\footnote{Allowing for mixed-strategy equilibria is attractive because finite games always have a mixed-strategy equilibrium. However, mixed equilibria are computationally demanding and may be less attractive for rationalizing stable market outcomes: realized pure actions can generate ex post incentives to deviate, whereas pure-strategy Nash equilibria do not have this drawback \citep{koh2023stable}.} Assumption \ref{assumption:independent.log.concave.unobservables} imposes conditional independence and log-concavity of the payoff shocks. These are standard distributional restrictions in empirical discrete-game models, and they are the key shock restrictions used below to obtain log-concave generalized likelihoods.

Assumption \ref{assumption:additive.separability.linear.payoffs} separates the deterministic payoff component from common and idiosyncratic unobservables. The common component $r_i(y_i)^\top S_i(\sigma)\lambda$ captures market-level unobserved heterogeneity that is observed by players but not by the econometrician. Because this component enters multiple players' payoffs, integrating out $\lambda$ induces correlation across players' composite payoff shocks, while conditional on $(x,\lambda)$ the idiosyncratic shocks remain independent. The formulation covers both unordered- and ordered-action games. In an unordered-action model with action-specific shocks, $r_i(y_i)$ can be a selector vector that picks out the shock associated with action $y_i$. In an ordered-action model with a scalar payoff shock entering proportionally to the action, one can take $r_i(y_i)=y_i$.

The key implication is that no-deviation payoff differences are affine in the payoff parameters and unobservables. For any profile $y$ and deviation $a_i\in\mathcal Y_i$,
\[
\begin{aligned}
D_{i,a_i}^{(\gamma,\sigma)}(y,x,\lambda,\varepsilon_i)
&\equiv
u_i^{(\gamma,\sigma)}(y_i,y_{-i},x,\lambda,\varepsilon_i)
-
u_i^{(\gamma,\sigma)}(a_i,y_{-i},x,\lambda,\varepsilon_i) \\
&=
v_i^\gamma(y_i,y_{-i},x)
-
v_i^\gamma(a_i,y_{-i},x) 
+
\left[
r_i(y_i)-r_i(a_i)
\right]^\top S_i(\sigma)\lambda
+
\left[
r_i(y_i)-r_i(a_i)
\right]^\top \varepsilon_i .
\end{aligned}
\]
Thus, for each fixed $\sigma$,
$D_{i,a_i}^{(\gamma,\sigma)}(y,x,\lambda,\varepsilon_i)$ is affine in
$(\gamma,\lambda,\varepsilon_i)$.

\subsection{Log-Concave Generalized Likelihoods}

This subsection establishes two properties of the singleton-event generalized likelihood. First, it admits a single-agent best-response representation: conditional on the common shock, the probability that a candidate profile is an equilibrium factors into the product of players' best-response probabilities. Second, under the primitive assumptions above, this likelihood is log-concave in the payoff parameters for each fixed value of the common-shock parameter.

Fix a candidate action profile $y\in\mathcal Y$. The singleton-event generalized likelihood conditional on the common shock is
\[
\mathcal L_{(\gamma,\sigma)}(y\mid x,\lambda)
\equiv
\Pr_{\varepsilon}\left(
\left\{
\varepsilon:
y\in G_{(\gamma,\sigma)}(x,\lambda,\varepsilon)
\right\}
\mid x,\lambda
\right),
\]
where $\varepsilon=(\varepsilon_1,\ldots,\varepsilon_I)$ and the probability is taken over idiosyncratic shocks. The corresponding unconditional generalized likelihood is
\[
\mathcal L_{(\gamma,\sigma)}(y\mid x)
\equiv
\Pr_{\varepsilon,\lambda}\left(
\left\{
(\varepsilon,\lambda):
y\in G_{(\gamma,\sigma)}(x,\lambda,\varepsilon)
\right\}
\mid x
\right)
=
\int
\mathcal L_{(\gamma,\sigma)}(y\mid x,\lambda)h(\lambda)\,d\lambda,
\]
where $h$ denotes the density of the common shock. 

Define the player-level best-response probability as
\[
\mathcal M_i^{(\gamma,\sigma)}(y_i\mid y_{-i},x,\lambda)
\equiv
\Pr_{\varepsilon_i}\left(
\left\{
\varepsilon_i:
D_{i,a_i}^{(\gamma,\sigma)}(y,x,\lambda,\varepsilon_i)\geq 0
\text{ for all } a_i\in\mathcal Y_i
\right\}
\mid x,\lambda
\right).
\]
Thus, $\mathcal M_i^{(\gamma,\sigma)}(y_i\mid y_{-i},x,\lambda)$ is the single-agent probability that $y_i$ is optimal when opponents' actions are fixed at $y_{-i}$.

\begin{theorem}[Product representation and log-concavity]
\label{theorem:primitive.log.concavity.singleton}
Suppose Assumptions~\ref{assumption:psne},
\ref{assumption:independent.log.concave.unobservables}, and
\ref{assumption:additive.separability.linear.payoffs} hold. Then, for each fixed $\sigma\in\Sigma$, each $y\in\mathcal Y$, and each $x\in\mathcal X$,
\[
\mathcal L_{(\gamma,\sigma)}(y\mid x,\lambda)
=
\prod_{i=1}^I
\mathcal M_i^{(\gamma,\sigma)}(y_i\mid y_{-i},x,\lambda),
\]
and
\[
\mathcal L_{(\gamma,\sigma)}(y\mid x)
=
\int
\prod_{i=1}^I
\mathcal M_i^{(\gamma,\sigma)}(y_i\mid y_{-i},x,\lambda)
h(\lambda)\,d\lambda.
\]
Moreover, $\mathcal M_i^{(\gamma,\sigma)}(y_i\mid y_{-i},x,\lambda)$ is log-concave in $(\gamma,\lambda)$,
$\mathcal L_{(\gamma,\sigma)}(y\mid x,\lambda)$ is log-concave in $(\gamma,\lambda)$, and
$\mathcal L_{(\gamma,\sigma)}(y\mid x)$ is log-concave in $\gamma$.
\end{theorem}

\begin{proof}
See Appendix~\ref{section:proof.of.theorem.2}.
\end{proof}

The product representation follows directly from the definition of pure-strategy Nash equilibrium. Conditional on $(x,\lambda)$ and a candidate profile $y$,
\[
\begin{split}
\left\{
\varepsilon:
y\in G_{(\gamma,\sigma)}(x,\lambda,\varepsilon)
\right\}
&=
\bigcap_{i=1}^I
\left\{
\varepsilon:
D_{i,a_i}^{(\gamma,\sigma)}(y,x,\lambda,\varepsilon_i)\geq 0
\text{ for all } a_i\in\mathcal Y_i
\right\} \\
&=
\times_{i=1}^I
\left\{
\varepsilon_i:
D_{i,a_i}^{(\gamma,\sigma)}(y,x,\lambda,\varepsilon_i)\geq 0
\text{ for all } a_i\in\mathcal Y_i
\right\}.
\end{split}
\]
The first equality states that all players must find their proposed actions optimal against the candidate profile. The second uses the fact that, conditional on $(x,\lambda)$ and $y$, player $i$'s best-response event depends only on $\varepsilon_i$. Conditional independence of the idiosyncratic shocks therefore gives the product of the player-level probabilities. The common shock is integrated out only after this conditioning step, producing an integral of the product rather than a product of unconditional probabilities.

The log-concavity argument is similarly direct. For fixed $\sigma$, define the best-response region
\[
B_i^\sigma(y,x)
\equiv
\left\{
(\gamma,\lambda,\varepsilon_i):
D_{i,a_i}^{(\gamma,\sigma)}(y,x,\lambda,\varepsilon_i)\geq 0
\text{ for all } a_i\in\mathcal Y_i
\right\}.
\]
By Assumption~\ref{assumption:additive.separability.linear.payoffs}, each payoff difference $D_{i,a_i}^{(\gamma,\sigma)}$ is affine in $(\gamma,\lambda,\varepsilon_i)$ for fixed $\sigma$, so $B_i^\sigma(y,x)$ is convex. Its indicator, multiplied by the log-concave density of $\varepsilon_i$, is therefore log-concave in $(\gamma,\lambda,\varepsilon_i)$. Pr\'ekopa's theorem implies that $\mathcal M_i^{(\gamma,\sigma)}$ is log-concave in $(\gamma,\lambda)$. Products preserve log-concavity; multiplying by the log-concave density $h(\lambda)$ and integrating out $\lambda$ preserves log-concavity again by Pr\'ekopa's theorem.

Theorem~\ref{theorem:primitive.log.concavity.singleton} is the main convexity result. If $\Gamma$ is convex, then each fixed-$\sigma$ slice $\Gamma_I^*(\sigma)$ is convex. Indeed, the singleton-event restrictions in \eqref{equation:gamma.slice} can be written as
\[
\log\phi(y\mid x) - \log\mathcal L_{(\gamma,\sigma)}(y\mid x) \leq 0,
\qquad
y\in\mathcal Y,\ x\in\mathcal X,
\]
with the convention that the restriction is vacuous when $\phi(y\mid x)=0$. Since $\log\mathcal L_{(\gamma,\sigma)}(y\mid x)$ is concave in $\gamma$, each restriction is convex in $\gamma$ for fixed $\sigma$. Thus the singleton-event outer set is a union of convex slices, as in \eqref{equation:outer.set.is.union.of.convex.sets}.

The computational implication is that projection bounds can be obtained by convex optimization. For each fixed value of the common-shock parameter $\sigma$, one optimizes linear functionals of $\gamma$ subject to the convex singleton-event restrictions, replacing pointwise membership testing over a high-dimensional parameter space with convex projection problems indexed by a low-dimensional nuisance parameter. The result is specific to singleton events; non-singleton event classes generally require separate arguments because unions of convex equilibrium regions need not be convex.

%% file: S.04.Unordered.Actions.tex
\section{Unordered Actions}
\label{section:unordered.actions}

This section specializes the singleton-event construction to games with unordered, multinomial actions. The primitive result in Section~\ref{section:convex.outer.set} establishes log-concavity under affine payoff differences and log-concave shocks. The Type-I extreme-value assumption adds a computational benefit: it gives the player-level best-response probabilities closed-form multinomial-logit expressions. Thus, in unordered-action games, the singleton-event generalized likelihood is available in closed form up to the low-dimensional integration over the common shock.

\subsection{Closed-Form Singleton-Event Likelihoods}

Let $\varepsilon_i=(\varepsilon_i(a_i))_{a_i\in\mathcal Y_i}$ denote player $i$'s vector of action-specific idiosyncratic payoff shocks. Let $r_i(a_i)$ be the selector vector that picks out the component of $\varepsilon_i$ associated with action $a_i$. Then the payoff specification in Assumption~\ref{assumption:additive.separability.linear.payoffs} can be written as
\[
u_i^{(\gamma,\sigma)}(a_i,y_{-i},x,\lambda,\varepsilon_i)
=
V_i^{(\gamma,\sigma)}(a_i,y_{-i},x,\lambda)
+
\varepsilon_i(a_i),
\]
where
\[
V_i^{(\gamma,\sigma)}(a_i,y_{-i},x,\lambda)
\equiv
v_i^\gamma(a_i,y_{-i},x)
+
r_i(a_i)^\top S_i(\sigma)\lambda .
\]
For fixed $\sigma$, $V_i^{(\gamma,\sigma)}$ is affine in $(\gamma,\lambda)$.

\begin{assumption}[Type-I extreme-value shocks for unordered actions]
\label{assumption:type.1.extreme.value.unordered}
The action-specific idiosyncratic payoff shocks
$\{\varepsilon_i(a_i): i=1,\ldots,I,\ a_i\in\mathcal Y_i\}$ are independent of $(x,\lambda)$, mutually independent across players and actions, and identically distributed Type-I extreme value.
\end{assumption}

\begin{theorem}[Unordered logit specialization]
\label{theorem:unordered.singleton.outer.set}
Suppose Assumptions~\ref{assumption:psne},
\ref{assumption:independent.log.concave.unobservables},
\ref{assumption:additive.separability.linear.payoffs}, and
\ref{assumption:type.1.extreme.value.unordered} hold. Then, for each player $i$, action profile $y$, covariate value $x$, and common shock $\lambda$,
\[
\mathcal M_i^{(\gamma,\sigma)}(y_i\mid y_{-i},x,\lambda)
=
\frac{
\exp\left(
V_i^{(\gamma,\sigma)}(y_i,y_{-i},x,\lambda)
\right)
}{
\sum_{a_i\in\mathcal Y_i}
\exp\left(
V_i^{(\gamma,\sigma)}(a_i,y_{-i},x,\lambda)
\right)
}.
\]
Consequently,
\[
\mathcal L_{(\gamma,\sigma)}(y\mid x,\lambda)
=
\prod_{i=1}^I
\frac{
\exp\left(
V_i^{(\gamma,\sigma)}(y_i,y_{-i},x,\lambda)
\right)
}{
\sum_{a_i\in\mathcal Y_i}
\exp\left(
V_i^{(\gamma,\sigma)}(a_i,y_{-i},x,\lambda)
\right)
},
\]
and
\[
\mathcal L_{(\gamma,\sigma)}(y\mid x)
=
\int
\prod_{i=1}^I
\frac{
\exp\left(
V_i^{(\gamma,\sigma)}(y_i,y_{-i},x,\lambda)
\right)
}{
\sum_{a_i\in\mathcal Y_i}
\exp\left(
V_i^{(\gamma,\sigma)}(a_i,y_{-i},x,\lambda)
\right)
}
h(\lambda)\,d\lambda .
\]
For each fixed $\sigma\in\Sigma$, $\mathcal L_{(\gamma,\sigma)}(y\mid x)$ is log-concave in $\gamma$. Therefore, if $\Gamma$ is convex, each slice $\Gamma_I^{*}(\sigma)$ is convex.
\end{theorem}

\begin{proof}
See Appendix~\ref{section:proof.of.theorem.3}.
\end{proof}

The theorem gives the main computational benefit of the unordered logit specification. Conditional on the common shock $\lambda$, evaluating the singleton-event generalized likelihood requires only the product of multinomial-logit best-response probabilities. It does not require simulation over idiosyncratic payoff shocks or equilibrium enumeration over simulated shock draws. Integrating over $\lambda$ then gives the unconditional generalized likelihood used in the singleton-event outer set.

This specialization is particularly relevant for binary entry games, which are among the most common applications of complete-information discrete games. With binary actions, the unordered and ordered representations coincide up to notation; the distinction becomes substantive only with richer action spaces. The singleton-event outer set can be used either as an informative object in its own right or as a fast first-stage screen before applying sharper but more expensive procedures. Section~\ref{section:empirical.application} provides an empirical illustration.

The same logic extends to vector-valued discrete actions. One can treat each action vector as a single multinomial alternative and apply the singleton-event construction directly. When the vector action space is large, coordinate-deviation restrictions provide a more tractable outer-set approximation; under additional structure ensuring that coordinatewise optimality characterizes full best responses, this approximation coincides with the singleton-event generalized likelihood. Appendix~\ref{section:extension.to.vector.decisions} develops this extension and connects it to product-entry settings related to \citet{fan2025estimating}.

The Type-I extreme-value assumption is stronger than what is needed for convexity. The primitive theorem in Section~\ref{section:convex.outer.set} only requires log-concavity of the relevant shock density and affine no-deviation payoff differences. What logit adds is closed-form evaluation: it delivers the expression in Theorem~\ref{theorem:unordered.singleton.outer.set} and avoids numerical integration over idiosyncratic payoff shocks. Logit should therefore be viewed as a tractable specification that delivers both convexity and closed-form best-response probabilities, not as a knife-edge assumption required for convexity itself.

\subsection{Illustration and Identifying Power}

I illustrate the singleton-event restrictions using the running two-player entry example without the common shock. In this case, $\theta=(\beta_1,\beta_2,\Delta_1,\Delta_2)$ belongs to $\Theta_I^{*}$ if and only if, for all $x\in\mathcal X$,
\begin{equation}
\label{equation:singleton.inequalities.example}
\begin{aligned}
\phi((0,0)\mid x) &\leq F(\underline{\epsilon}_1(x))F(\underline{\epsilon}_2(x)),\\
\phi((1,1)\mid x) &\leq \bigl(1-F(\overline{\epsilon}_1(x))\bigr) \bigl(1-F(\overline{\epsilon}_2(x))\bigr),\\
\phi((1,0)\mid x) &\leq \bigl(1-F(\underline{\epsilon}_1(x))\bigr) F(\overline{\epsilon}_2(x)),\\
\phi((0,1)\mid x) &\leq F(\overline{\epsilon}_1(x)) \bigl(1-F(\underline{\epsilon}_2(x))\bigr),
\end{aligned}
\end{equation}
where $\underline{\epsilon}_i(x)=-x_i^\top\beta_i$, $\overline{\epsilon}_i(x)=\underline{\epsilon}_i(x)-\Delta_i$,  and $F(z)=1/(1+e^{-z})$ is the standard logistic cumulative distribution function.\footnote{The standard logistic distribution arises because, in the binary-action logit specification, the payoff-relevant scalar shock can be written as the difference between two independent Type-I extreme-value action-specific shocks. If $\varepsilon_i(1)$ and $\varepsilon_i(0)$ are independent Type-I extreme-value shocks with the standard scale normalization, then $\varepsilon_i(1)-\varepsilon_i(0)$ follows the standard logistic distribution.}

These inequalities illustrate both the computational and identifying content of the singleton-event outer set. Computationally, singleton events correspond to rectangular best-response regions in the payoff-shock space, whose probabilities are easy to evaluate under logit shocks. Substantively, the inequalities rule out extreme parameter values. For example, if $\phi((1,1)\mid x)$ is bounded away from zero, then the competitive effects cannot make joint entry nearly impossible, because the second inequality in \eqref{equation:singleton.inequalities.example} would be violated.

A simple numerical example suggests that the singleton-event outer set can
nevertheless be quite tight. Consider the two-player entry game with payoff $u_i^\theta(y_i,y_{-i}) = y_i(\beta_i+\Delta_i y_{-i}+\epsilon_i)$, with $\beta_i=0$, $\Delta_i=-0.5$, i.i.d.\ standard logistic $\epsilon_i$, and a symmetric equilibrium selection rule. The model then implies the choice-probability vector $(\phi_{00},\phi_{10},\phi_{01},\phi_{11})
\approx 
(0.250,\,0.304,\,0.304,\,0.143).$

\begin{figure}[htbp!]
\centering

\begin{subfigure}[t]{0.47\textwidth}
    \centering
    \includegraphics[width=\textwidth]{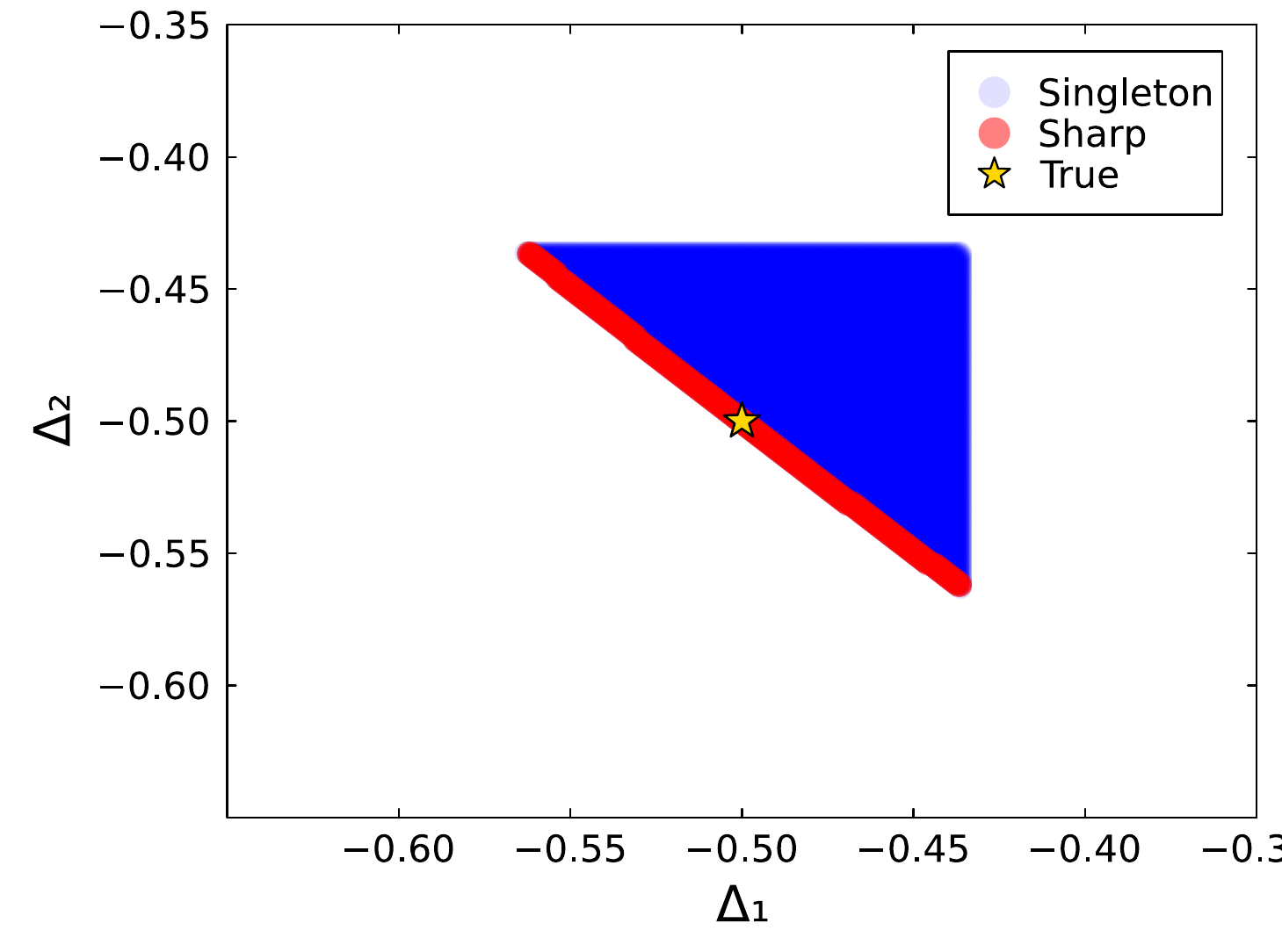}
    \caption{$(\Delta_1,\Delta_2)$ with $(\beta_1,\beta_2)$ fixed}
\end{subfigure}
\hfill
\begin{subfigure}[t]{0.47\textwidth}
    \centering
    \includegraphics[width=\textwidth]{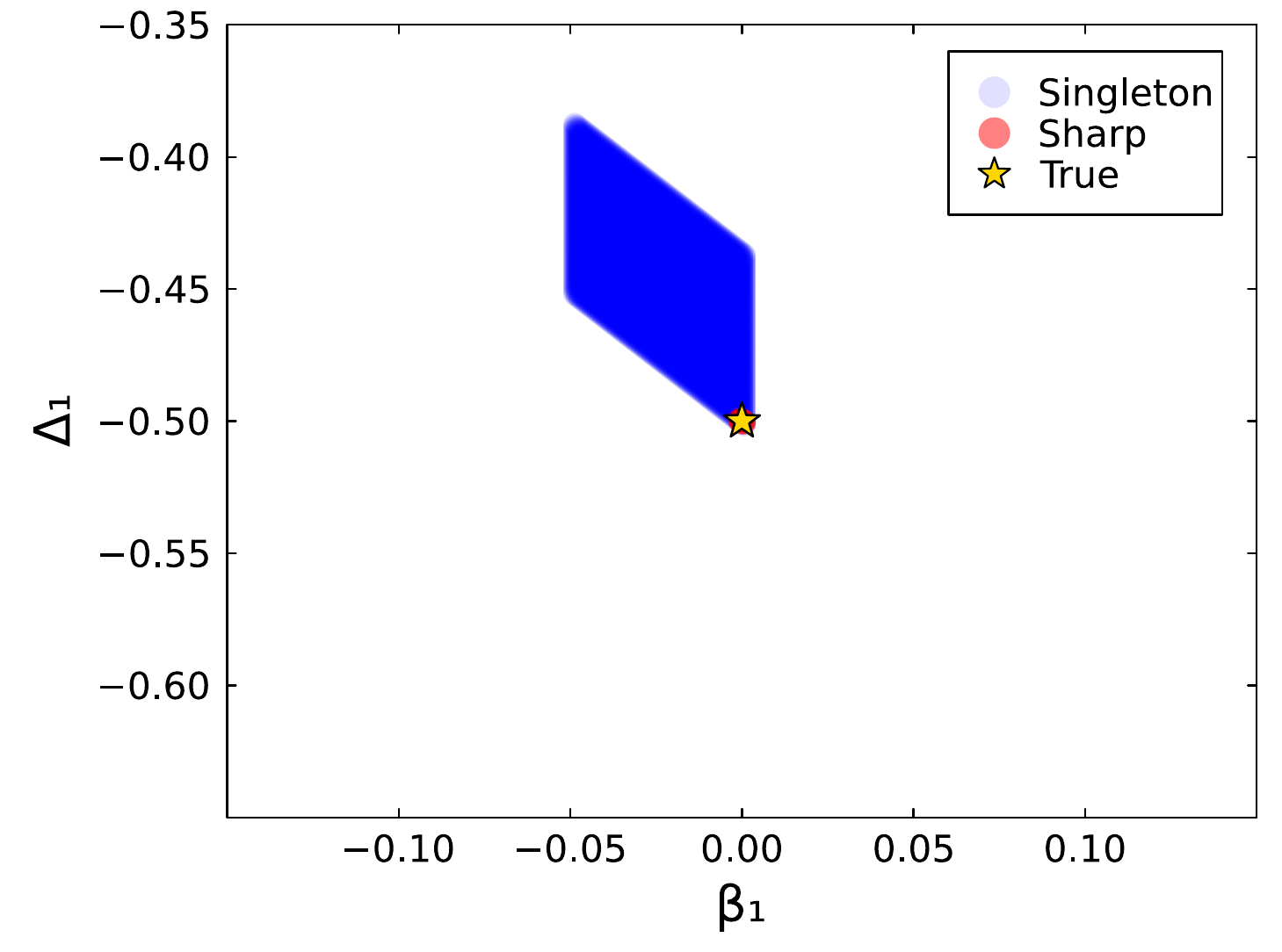}
    \caption{$(\beta_1,\Delta_1)$ with $(\beta_2,\Delta_2)$ fixed}
\end{subfigure}

\vspace{0.5em}

\begin{subfigure}[t]{0.47\textwidth}
    \centering
    \includegraphics[width=\textwidth]{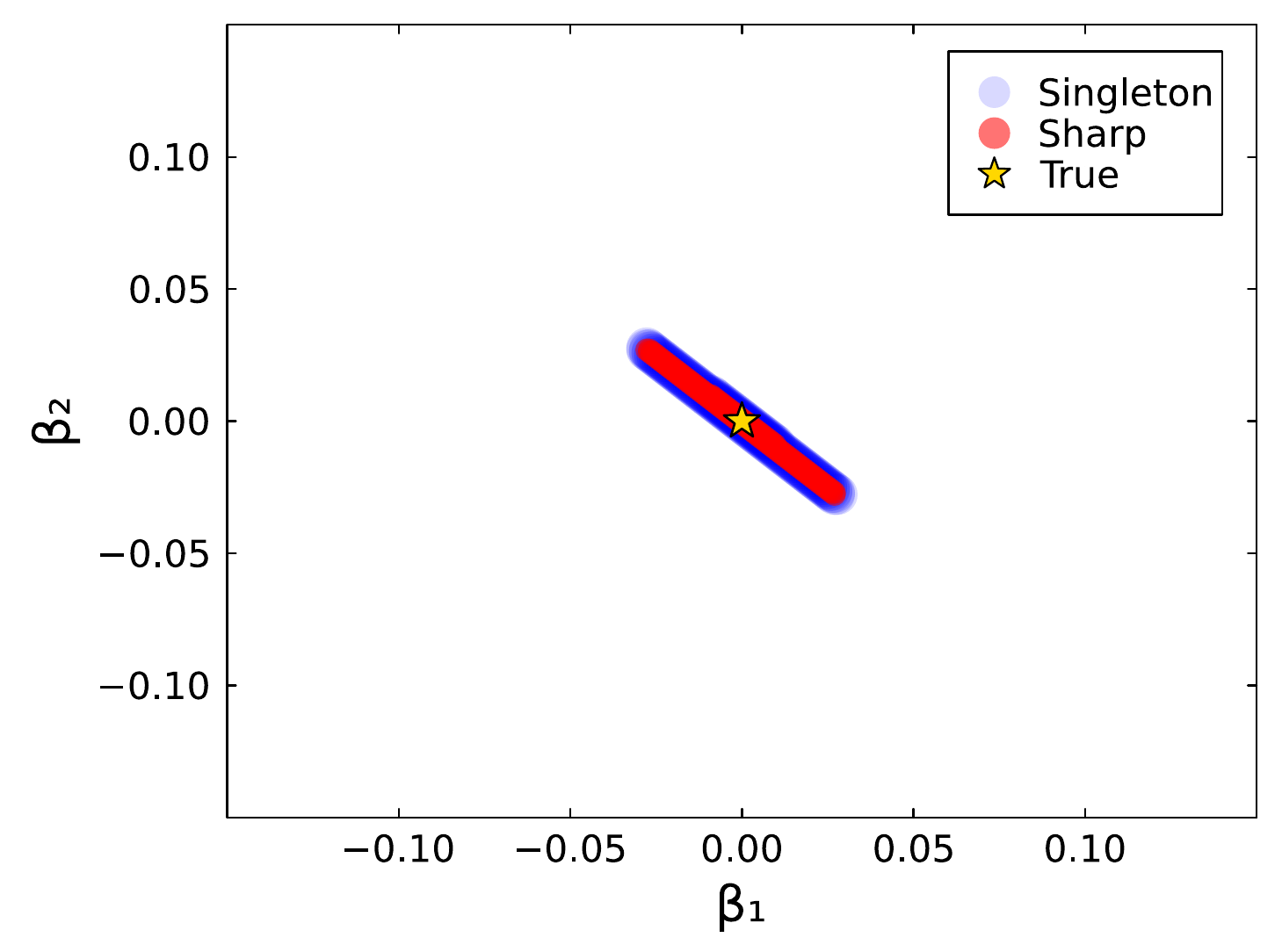}
    \caption{$(\beta_1,\beta_2)$ with $(\Delta_1,\Delta_2)$ fixed}
\end{subfigure}

\caption{Two-dimensional slices of the singleton-event outer set and sharp identified set}
\label{fig:singleton_three_slices}
\begin{minipage}{0.95\textwidth}
\footnotesize
\emph{Notes}: The figure compares the singleton-event outer set (blue) and the sharp identified set (red) in three two-dimensional slices of the parameter space. The star denotes the true parameter. Panel (a) fixes $(\beta_1,\beta_2)$ and varies $(\Delta_1,\Delta_2)$; panel (b) fixes $(\beta_2,\Delta_2)$ and varies $(\beta_1,\Delta_1)$; and panel (c) fixes $(\Delta_1,\Delta_2)$ and varies $(\beta_1,\beta_2)$. In panel (b), the sharp set is visually indistinguishable from the true parameter, as it is a singleton that coincides with it. In panel (c), the singleton-event outer set coincides with the sharp set; the singleton-event points are plotted with larger markers than the sharp-set points to make this overlap visible. In all panels, the singleton-event outer set contains the sharp set.
\end{minipage}
\end{figure}

Figure \ref{fig:singleton_three_slices} compares the singleton-event outer set with the sharp set in three two-dimensional slices of the parameter space. Panel (a) fixes $(\beta_1,\beta_2)$ at their true values and searches over $(\Delta_1,\Delta_2)$. Panel (b) fixes $(\beta_2,\Delta_2)$ at their true values and searches over $(\beta_1,\Delta_1)$. Panel (c) fixes $(\Delta_1,\Delta_2)$ at their true values and searches over $(\beta_1,\beta_2)$. In all three panels, the singleton-event outer set contains the sharp set and the true parameter. The gap between the two sets is visible in panels (a) and (b), especially when the sharp set is highly localized, but the singleton-event restrictions still substantially restrict the parameter space. In panel (c), once $(\Delta_1,\Delta_2)$ are fixed, the singleton-event outer set is visually nearly indistinguishable from the sharp set.\footnote{The axes report raw payoff coefficients under the standard logistic shock normalization. They are not rescaled by the shock standard deviation $\pi/\sqrt{3}\approx 1.81$. The apparent convexity of the sharp set in these slices is a feature of this numerical example and should not be interpreted as a general property.} 

Taken together, these figures suggest that singleton-event inequalities retain substantial identifying power even though they do not deliver the sharp identified set. They therefore provide a computationally simple outer approximation that can be informative in practice. 

Table \ref{table:unordered.projection.intervals} reports projection intervals computed by allowing all four parameters $(\beta_1,\beta_2,\Delta_1,\Delta_2)$ to vary jointly, without imposing symmetry across players. The singleton-event and sharp projection intervals are similar, consistent with the visual evidence in Figure \ref{fig:singleton_three_slices}. This further suggests that the singleton-event restrictions preserve substantial identifying power despite not being sharp.

\begin{table}[htbp!]
\small
\caption{Comparison of projection intervals}
\label{table:unordered.projection.intervals}
\centering
\begin{threeparttable}
\begin{tabular}{lcccc}
\toprule
 & $\beta_{1}$ & $\beta_{2}$ & $\Delta_{1}$ & $\Delta_{2}$ \\
\midrule
Sharp
& $[-0.193,0.181]$
& $[-0.199,0.174]$
& $[-0.910,-0.055]$
& $[-0.902,-0.054]$ \\ [0.3ex]

Singleton-event
& $[-0.199,0.184]$
& $[-0.216,0.181]$
& $[-0.939,-0.030]$
& $[-0.921,-0.007]$ \\
\bottomrule
\end{tabular}
\begin{tablenotes}
\footnotesize
\item \emph{Notes}: Intervals are grid approximations to the population projection bounds over $(\beta_1,\beta_2,\Delta_1,\Delta_2)$, computed without imposing symmetry across players. The singleton-event outer set delivers projection bounds close to those of the sharp identified set, suggesting that the singleton-event restrictions retain substantial identifying power.
\end{tablenotes}
\end{threeparttable}
\end{table}

%% file: S.05.OrderedActions.tex
\section{Ordered Actions}
\label{section:ordered.actions}

This section specializes the singleton-event construction to ordered-action games in the framework of \citet{aradillas2022inference}. The primitive result in Section~\ref{section:convex.outer.set} establishes log-concavity under affine payoff differences and log-concave shocks. The ordered-action structure adds a computational benefit: best responses can be represented by threshold-crossing rules. With logistic scalar payoff shocks, the player-level best-response probabilities take ordered-logit form, so the singleton-event generalized likelihood is available in closed form up to the low-dimensional integration over the common shock.\footnote{Relative to \citet{aradillas2022inference}, who use \citet{chesher2017generalized}'s characterization of the sharp identified set, I use \citet{galichon2011setidentification}'s characterization and provide additional results for pure-strategy Nash equilibria.}

\subsection{Setup and Threshold Representation}

Each player chooses from a finite ordered action set $\mathcal Y_i=\{s_i^0,s_i^1,\ldots,s_i^{L_i}\}\subseteq\mathbb R$, where $s_i^0<s_i^1<\cdots<s_i^{L_i}$.\footnote{It is common to take $s_i^\ell=\ell$, so that $\mathcal Y_i=\{0,1,\ldots,L_i\}$.} If $y_i=s_i^\ell$, let $y_i^+=s_i^{\ell+1}$ denote the adjacent higher action, with the convention that $s_i^{L_i+1}$ is used only as a boundary label.

In the ordered-action case, I specialize
Assumption~\ref{assumption:additive.separability.linear.payoffs} by taking the
shock loading to be $r_i(y_i)=y_i$. Thus,
\[
u_i^{(\gamma,\sigma)}(y,x,\lambda,\varepsilon_i)
=
V_i^{(\gamma,\sigma)}(y,x,\lambda)+y_i\varepsilon_i,
\]
where
\[
V_i^{(\gamma,\sigma)}(y,x,\lambda)
\equiv
v_i^\gamma(y,x)+y_iS_i(\sigma)\lambda .
\]
For each fixed $\sigma$, $V_i^{(\gamma,\sigma)}$ is affine in
$(\gamma,\lambda)$. The scalar shock $\varepsilon_i$ shifts the payoff toward
higher actions because it enters proportionally to $y_i$.

\begin{assumption}[Strict concavity in own action]
\label{assumption:strict.concavity}
For each player $i$, opponents' action profile $y_{-i}$, covariate value $x$,
common shock $\lambda$, and parameter value $(\gamma,\sigma)$, the deterministic
payoff component $V_i^{(\gamma,\sigma)}(\cdot,y_{-i},x,\lambda)$ is strictly
concave on the ordered action grid. That is, for every interior action
$s_i^\ell$, $\ell=1,\ldots,L_i-1$, $\frac{
V_i^{(\gamma,\sigma)}(s_i^{\ell+1},y_{-i},x,\lambda)
-
V_i^{(\gamma,\sigma)}(s_i^\ell,y_{-i},x,\lambda)
}{
s_i^{\ell+1}-s_i^\ell
}
<
\frac{
V_i^{(\gamma,\sigma)}(s_i^\ell,y_{-i},x,\lambda)
-
V_i^{(\gamma,\sigma)}(s_i^{\ell-1},y_{-i},x,\lambda)
}{
s_i^\ell-s_i^{\ell-1}
}.$
\end{assumption}

Strict concavity is the only shape restriction needed for the threshold representation below.\footnote{Since the idiosyncratic shock enters as $y_i\varepsilon_i$, increasing differences in $(y_i,\varepsilon_i)$ hold automatically. A decreasing-differences assumption in $(y_i,y_{-i})$ would imply that thresholds are weakly increasing in opponents' actions, but that comparative-static restriction is not needed for the likelihood or convexity result.}  For $\ell=1,\ldots,L_i$, define the adjacent-action threshold
\[
e_i^{(\gamma,\sigma)}(s_i^\ell,y_{-i},x,\lambda)
\equiv
\frac{
V_i^{(\gamma,\sigma)}(s_i^{\ell-1},y_{-i},x,\lambda) - V_i^{(\gamma,\sigma)}(s_i^\ell,y_{-i},x,\lambda)
}{
s_i^\ell-s_i^{\ell-1}
}.
\]
Set $e_i^{(\gamma,\sigma)}(s_i^0,y_{-i},x,\lambda)\equiv -\infty,$ and $e_i^{(\gamma,\sigma)}(s_i^{L_i+1},y_{-i},x,\lambda)\equiv +\infty$.

\begin{lemma}[Threshold representation]
\label{lemma:threshold.crossing.rule}
Suppose the ordered-action payoff specification above holds and Assumption~\ref{assumption:strict.concavity} holds. Then, for each player $i$, opponents' action profile $y_{-i}$, covariate value $x$, and common shock $\lambda$, the thresholds are strictly ordered:
\[
-\infty = e_i^{(\gamma,\sigma)}(s_i^0,y_{-i},x,\lambda) < e_i^{(\gamma,\sigma)}(s_i^1,y_{-i},x,\lambda) < \cdots < e_i^{(\gamma,\sigma)}(s_i^{L_i},y_{-i},x,\lambda) < e_i^{(\gamma,\sigma)}(s_i^{L_i+1},y_{-i},x,\lambda) = +\infty .
\]
Moreover, for any fixed parameter value $(\gamma,\sigma)$, state
$(x,\lambda,\varepsilon)$, and action profile $y$, $y$ is a pure-strategy Nash
equilibrium if and only if, for every player $i$,
\[
e_i^{(\gamma,\sigma)}(y_i,y_{-i},x,\lambda) \leq \varepsilon_i \leq e_i^{(\gamma,\sigma)}(y_i^+,y_{-i},x,\lambda).
\]
When the relevant finite inequalities are strict, $y_i$ is player $i$'s unique
best response to $y_{-i}$.
\end{lemma}

\begin{proof}
See Appendix~\ref{section:proof.of.lemma.threshold.crossing.rule}.
\end{proof}

The lemma shows that, conditional on $(x,\lambda)$ and opponents' actions $y_{-i}$, player $i$'s best-response event is an interval restriction on the scalar shock $\varepsilon_i$. Therefore, the player-level best-response probability $\mathcal M_i^{(\gamma,\sigma)}(y_i\mid y_{-i},x,\lambda)$ is an interval probability. Appendix~\ref{appendix:ordered.action.illustrations} illustrates the threshold construction in the binary-action case.

\subsection{Closed-Form Singleton-Event Likelihoods}

I now impose a logistic distribution on the scalar idiosyncratic payoff shocks
to obtain closed-form player-level best-response probabilities.

\begin{assumption}[Logistic scalar payoff shocks]
\label{assumption:logistic.distribution.ordered}
The scalar idiosyncratic payoff shocks
$\varepsilon_1,\ldots,\varepsilon_I$ are independent of $(x,\lambda)$,
mutually independent across players, and identically distributed according to
the standard logistic distribution. Let
$F(t)=1/(1+\exp(-t))$ denote the standard logistic cumulative distribution
function.
\end{assumption}

\begin{theorem}[Ordered logit specialization]
\label{theorem:ordered.singleton.outer.set}
Suppose Assumptions~\ref{assumption:psne},
\ref{assumption:independent.log.concave.unobservables},
\ref{assumption:additive.separability.linear.payoffs},
\ref{assumption:strict.concavity}, and
\ref{assumption:logistic.distribution.ordered} hold. Then, for each player
$i$, action profile $y$, covariate value $x$, and common shock $\lambda$,
\[
\mathcal M_i^{(\gamma,\sigma)}(y_i\mid y_{-i},x,\lambda)
=
F\!\left(
e_i^{(\gamma,\sigma)}(y_i^+,y_{-i},x,\lambda)
\right)
-
F\!\left(
e_i^{(\gamma,\sigma)}(y_i,y_{-i},x,\lambda)
\right).
\]
Consequently,
\[
\mathcal L_{(\gamma,\sigma)}(y\mid x,\lambda)
=
\prod_{i=1}^I
\left[
F\!\left(
e_i^{(\gamma,\sigma)}(y_i^+,y_{-i},x,\lambda)
\right)
-
F\!\left(
e_i^{(\gamma,\sigma)}(y_i,y_{-i},x,\lambda)
\right)
\right],
\]
and
\[
\mathcal L_{(\gamma,\sigma)}(y\mid x)
=
\int
\prod_{i=1}^I
\left[
F\!\left(
e_i^{(\gamma,\sigma)}(y_i^+,y_{-i},x,\lambda)
\right)
-
F\!\left(
e_i^{(\gamma,\sigma)}(y_i,y_{-i},x,\lambda)
\right)
\right]
h(\lambda)\,d\lambda .
\]
For each fixed $\sigma\in\Sigma$,
$\mathcal L_{(\gamma,\sigma)}(y\mid x)$ is log-concave in $\gamma$.
Therefore, if $\Gamma$ is convex, each slice
$\Gamma_I^{*}(\sigma)$ is convex.
\end{theorem}

\begin{proof}
See Appendix~\ref{section:proof.of.theorem.ordered.singleton.outer.set}.
\end{proof}

The theorem gives the ordered-action analogue of the multinomial-logit formula
in Theorem~\ref{theorem:unordered.singleton.outer.set}. Conditional on the
common shock $\lambda$, evaluating the singleton-event generalized likelihood
requires only the product of ordered-logit interval probabilities. It does not
require simulation over idiosyncratic payoff shocks or equilibrium enumeration
over simulated shock draws. Integrating over $\lambda$ then gives the
unconditional generalized likelihood used in the singleton-event outer
set.\footnote{Although the ordered-logit probabilities are written as
differences of CDF values, $F(\overline e)-F(\underline e)$, the convexity
argument treats them as interval probabilities
$\Pr(\underline e\leq \varepsilon_i\leq \overline e)$ over regions defined by
affine inequalities. Log-concavity follows from the primitive
Pr\'ekopa argument, not from a general algebraic property of CDF differences.} Appendix \ref{appendix:ordered.action.illustrations} provides a numerical illustration based on a simple two-player multi-store entry game.

The ordered-action specialization shows that the singleton-event construction
is not limited to unordered or binary-action games. The same object
$\Theta_I^{*}$ applies: the only difference is how the player-level
best-response probability $\mathcal M_i$ is evaluated. In unordered-action
logit games, $\mathcal M_i$ is a multinomial-logit probability. In
ordered-action logit games, $\mathcal M_i$ is an ordered-logit interval
probability. In both cases, the observed CCP of a profile is bounded above by
the probability that the profile passes all players' single-agent
best-response tests.

%% file: S.07.Empirical.Application.tex
\section{Empirical Application I: Discount-Store Entry Game\label{section:empirical.application}}

I apply the proposed method to the Walmart--Kmart entry game studied by 
\citet{ellickson_structural_2011}, which adapts \citet{jia_what_2008}'s model to independent county-level entry games.\footnote{\citet{jia_what_2008} model Walmart and Kmart as playing a single network game across many markets, whereas \citet{ellickson_structural_2011} treat entry decisions as independent across local markets to illustrate standard discrete-game estimation methods that impose equilibrium-selection assumptions. I refer readers to these papers for details on the model and data.} The application provides a natural benchmark, as binary entry games are canonical in the literature; the results can be compared with estimates that impose equilibrium-selection assumptions; and the data are simple, transparent, and publicly available.

\subsection{Setup}

I model Walmart's and Kmart's county-level entry decisions as independent static binary-entry games. Each firm either operates a store or stays out, firms have complete information, and the solution concept is pure-strategy Nash equilibrium. Payoffs follow \eqref{equation:entry.game.utility.specification}, with a common non-positive competitive effect so that entry decisions are strategic substitutes. The sample contains 2,065 county markets.

\begin{table}[htbp!]
\small
\caption{Jia (2008) data summary statistics}
    \label{table:jia.data.summary.statistics}
    \centering
    \begin{threeparttable}
    \begin{tabular}{lccccc}
    \toprule
    Variable & Mean & Median & Std. Dev. & Min & Max \\ \midrule
         Walmart entry & 0.48 & 0.00 & 0.50 & 0.00 & 1.00  \\
         Kmart entry & 0.19 & 0.00 & 0.39 & 0.00 & 1.00 \\
         Population (thousand) & 2.98 & 2.98 & 0.67 & 1.54 & 4.37 \\
         Retail sales per capita (1984 \$, thousand) & 8.20 & 8.25 & 0.47 & 5.08 & 10.70 \\
         Percentage of urban population & 0.33 & 0.34 & 0.24 & 0.00 & 1.00 \\
         Distance to Benton, AR (100 miles) & 6.24 & 6.32 & 0.63 & 3.01 & 8.29 \\
         South region & 0.50 & 1.00 & 0.50 & 0.00 & 1.00 \\
         Midwest region & 0.42 & 0.00 & 0.49 & 0.00 & 1.00 \\ \midrule
         Number of markets (counties) & 2,065 \\ 
         \bottomrule
    \end{tabular}
    \begin{tablenotes}
    \footnotesize
        \item \emph{Notes}: Walmart entry and Kmart entry are market-level entry indicators. Population, retail sales per capita, urban population share, and distance to Benton are continuous market covariates. South equals one for markets in the Southwest or Southeast region; Midwest equals one for markets in the Great Lakes, Plains, or Rocky Mountain region.
    \end{tablenotes}
    \end{threeparttable}
\end{table}

Table \ref{table:jia.data.summary.statistics} summarizes the entry outcomes and covariates used in the application. To facilitate conditional-choice-probability estimation, I discretize the continuous variables---population, retail sales per capita, urban population share, and distance to Benton---into two support points using median splits and within-bin means.\footnote{Using auxiliary regressions of firms' entry decisions on covariates with and without discretization, I confirm that discretization does not materially change the regression coefficients.} After dropping covariate bins with no observations, I obtain $|\mathcal X|=40$ covariate bins.

\subsection{Estimation and Inference}

I construct confidence sets using the moment inequalities developed above. Let
$j=(A,x)$ index an event $A$ and a covariate cell $x$, let
$\phi_j\equiv \phi(A\mid x)$ denote the population conditional probability of
event $A$, and let $\mathcal L_{\theta,j}$ denote the corresponding
model-implied generalized-likelihood bound. For an event collection
$\mathcal J$, the population restrictions are
\[
    \phi_j \leq \mathcal L_{\theta,j},
    \qquad j\in\mathcal J .
\]

The first-stage object entering these restrictions is the vector of conditional
choice probabilities. To account for first-stage sampling error, I construct
simultaneous one-sided lower confidence bounds for the log event probabilities.
For each event-cell pair $j$, let
\[
    \ell_j
    =
    \log \widehat \phi_j
    -
    c_{1-\alpha}
    \frac{\widehat \tau_j}{\sqrt{n_x}},
    \qquad
    \widehat \tau_j
    =
    \left(\frac{1-\widehat \phi_j}{\widehat \phi_j}\right)^{1/2},
\]
where $c_{1-\alpha}$ is a simulated covariance-aware max-statistic critical
value. In the computation, I use regularized within-cell frequencies to avoid
degenerate log probabilities and use the same regularized first-stage
probabilities in the critical-value simulation, standard-error calculation, and
lower-bound construction. Appendix~\ref{section:computational.details}
describes these details. The resulting confidence set for event collection
$\mathcal J$ is
\[
    \widehat \Theta_I^{1-\alpha}(\mathcal J)
    =
    \left\{
    \theta \in \Theta:
    \ell_j \leq \log \mathcal L_{\theta,j},
    \quad j\in\mathcal J
    \right\}.
\]
This construction is conservative in the standard asymptotic sense: with
probability approaching at least $1-\alpha$, the lower envelope lies below the
true log event probabilities simultaneously, so any parameter vector satisfying
the population inequalities is retained, up to numerical tolerance.

I report two partial-identification confidence sets. The first uses the
singleton-event class, where $\mathcal J$ contains the four singleton action
profiles in each covariate cell. This gives a computationally tractable outer
confidence set. For each fixed value of the common-shock scale $\sigma$,
$\log \mathcal L_{\theta,j}$ is concave in the payoff parameters, so the
singleton-event feasible set is convex. The exact integrated restrictions imply convex projection problems over payoff
parameters conditional on the common-shock scale. In the implementation, I
approximate the one-dimensional common-shock integral by deterministic
quadrature and solve the resulting smooth projection problems over the retained
common-shock values. I first scan over a grid of $\sigma$ values and retain only those for
which the minimum numerical violation is below a fixed tolerance. I then compute
lower and upper projections for each payoff parameter at the retained grid
values of $\sigma$ and report the union of these projections. Payoff coefficients
are reported after normalizing by the standard deviation of the total payoff
shock, $\{\pi^2/3+\sigma^2\}^{1/2}$.

The second confidence set is a numerical approximation to the sharp confidence
set. It uses the same lower-bound construction, but recomputes the simultaneous
first-stage lower bounds for a core-determining collection of events. Because
the resulting projection problem is not solved by convex optimization, I use
the singleton-event outer confidence set to define a focused search region and
then search for parameter values satisfying the core inequalities. The reported
sharp-set intervals should therefore be interpreted as sampled numerical
approximations to the sharp confidence-set projections.

For comparison, I also estimate standard point-identified complete-information
benchmarks based on \citet{bresnahan_entry_1991} and
\citet{berry1992estimation}. The Bresnahan--Reiss benchmark treats firms
symmetrically and uses the number of entrants as the outcome. The Berry-style
benchmarks impose alternative equilibrium-selection or move-order assumptions:
one lets the more profitable firm move first, one lets Walmart move first, and
one lets Kmart move first. All benchmark models are estimated on the same
discretized covariate support used in the partial-identification exercise.
See Appendix~\ref{app:computation.first.application} for further computational details.

\subsection{Estimation Results}

Table \ref{table:empirical.results.for.the.two.chain.entry.application}
compares the partially identified estimates in Columns (5)--(6) with benchmark
point-identified estimates in Columns (1)--(4). Column (1) reports the
Bresnahan--Reiss specification. Columns (2)--(4) report Berry-style
specifications in which the moving firm is, respectively, the more profitable
firm, Walmart, or Kmart. Column (5) reports the singleton-event class outer
confidence set, and Column (6) reports the sampled sharp-set
approximation.\footnote{Column (6) reports ranges over accepted points found by
the numerical search. These ranges are inner numerical approximations and may be
narrower than the exact sharp-set projection intervals if the search misses
boundary points.}

All specifications use standard logistic idiosyncratic payoff shocks. The
partially identified specifications also allow for a common market-level payoff
shock, so that firm $i$'s unobserved payoff component is
$\sigma\lambda+\epsilon_i$, with $\lambda\sim N(0,1)$. The implied correlation
between the two firms' composite payoff shocks is $\rho(\sigma)=\frac{\sigma^2}{\sigma^2+\pi^2/3}.$
For the singleton-event class outer confidence set, I scan over a grid of
common-shock scales and retain the values that satisfy the feasibility
criterion.\footnote{The grid is
$\sigma\in\{0,0.5,1.0,\ldots,10.0\}$. For each grid value, I solve a
minimum-violation problem and retain the grid point if the minimized violation
is no larger than the numerical feasibility tolerance. The upper endpoint
$\sigma=10$ corresponds to
$\rho(10)=10^2/(10^2+\pi^2/3)\approx 0.968$.}
To make magnitudes comparable across specifications, coefficients are reported
in the normalized payoff scale: benchmark coefficients are divided by
$\sqrt{\pi^2/3}$, while each partially identified parameter value is divided by
$\sqrt{\pi^2/3+\sigma^2}$ using its associated common-shock scale.

\begin{table}[htbp!]
\scriptsize
\caption{Empirical results for the two-chain entry application}
\label{table:empirical.results.for.the.two.chain.entry.application}
\centering
\begin{threeparttable}
\begin{tabular}{lcccccc}
\toprule 
& \multicolumn{4}{c}{Benchmark point-identified models} 
& \multicolumn{2}{c}{Partially-identified models ($\alpha = 0.01$)} \\ 
\cmidrule(lr){2-5} \cmidrule(lr){6-7}
& (1) & (2) & (3) & (4) & (5) & (6) \\
& BR & Berry-Profit & Berry-Walmart & Berry-Kmart & Singleton-Class & Sharp Sample \\
\midrule
\hspace*{-0.5em}\emph{Common effects} \\
Population
& \(1.11\;(0.05)\)
& \(1.34\;(0.06)\)
& \(1.34\;(0.06)\)
& \(1.35\;(0.06)\)
& \([0.92,\;1.73]\)
& \([1.16,\;1.63]\) \\
Retail sales per capita
& \(0.91\;(0.07)\)
& \(1.16\;(0.08)\)
& \(1.15\;(0.08)\)
& \(1.16\;(0.08)\)
& \([-0.02,\;1.74]\)
& \([0.42,\;1.37]\) \\
Urban
& \(1.22\;(0.12)\)
& \(1.38\;(0.13)\)
& \(1.37\;(0.13)\)
& \(1.38\;(0.13)\)
& \([0.02,\;2.44]\)
& \([0.88,\;2.07]\) \\
Competitive effect \(\Delta\)
& \(-0.38\;(0.04)\)
& \(-0.12\;(0.05)\)
& \(-0.11\;(0.05)\)
& \(-0.12\;(0.05)\)
& \([-0.95,\;0.00]\)
& \([-0.86,\;-0.10]\) \\

\hspace*{-0.5em}\emph{Walmart-specific effects} \\
Intercept 
& \(-11.59\;(0.58)\)
& \(-8.91\;(0.78)\)
& \(-8.88\;(0.78)\)
& \(-8.93\;(0.78)\)
& \([-15.15,\;1.07]\)
& \([-12.35,\;-2.22]\) \\
Distance to Benton 
& --
& \(-0.86\;(0.08)\)
& \(-0.86\;(0.08)\)
& \(-0.87\;(0.08)\)
& \([-1.41,\;-0.26]\)
& \([-1.10,\;-0.40]\) \\
South 
& --
& \(0.61\;(0.07)\)
& \(0.61\;(0.07)\)
& \(0.61\;(0.07)\)
& \([-0.26,\;0.97]\)
& \([-0.11,\;0.62]\) \\

\hspace*{-0.5em}\emph{Kmart-specific effects} \\
Intercept 
& \(-11.59\;(0.58)\)
& \(-15.31\;(0.71)\)
& \(-15.28\;(0.71)\)
& \(-15.34\;(0.72)\)
& \([-19.53,\;-5.48]\)
& \([-16.87,\;-8.89]\) \\
Midwest 
& --
& \(0.22\;(0.07)\)
& \(0.22\;(0.07)\)
& \(0.22\;(0.07)\)
& \([-0.18,\;0.50]\)
& \([-0.10,\;0.33]\) \\
\midrule 
Accepted $\rho$ range 
& $-$ & $-$ & $-$ & $-$ & $[0.00, 0.93]$ & $[0.00, 0.86]$ \\
Runtime (sec.) 
& $-$ & $-$ & $-$ & $-$ & $112.3$ & $14060.9$ \\
\bottomrule
\end{tabular}
\begin{tablenotes}
\footnotesize
\item \emph{Notes}: 
Columns (1)--(4) report complete-information benchmark estimates on the same
discretized covariate support, with inverse-Hessian standard errors in
parentheses. Columns (5)--(6) report projection intervals for 99 percent confidence sets:
the singleton-event outer set and a sampled sharp-set approximation based on a
core-determining event collection. The accepted \(\rho\) range is the implied
range of composite-shock correlations,
\(\rho(\sigma)=\sigma^2/(\sigma^2+\pi^2/3)\), over accepted common-shock
scales. Parameters are reported in the normalized payoff scale, and
\(\Delta\leq 0\) is imposed. Runtime reports total computational time, not elapsed wall-clock time; for distributed runs, this is the sum of runtimes across worker processes. The sharp-set runtime excludes the singleton-class projection step used to construct its search region. 

\end{tablenotes}
\end{threeparttable}
\end{table}

The estimates deliver broadly consistent economic conclusions across
specifications. Market-size variables have positive effects on entry in the
complete-information benchmarks. Population remains positive throughout both
partial-identification confidence sets. Urban status is also positive throughout
both partial-identification intervals. Retail sales per capita is less precisely
bounded in the singleton-event class interval, which narrowly includes zero, but
is positive throughout the sampled sharp-set interval. The competitive effect is
negative in all benchmark specifications. The singleton-event class interval
reaches zero only at the boundary imposed by the nonpositivity restriction on
\(\Delta\), while the sampled sharp-set interval is bounded away from zero.

The firm-specific estimates are also intuitive. Distance to Benton has a
negative effect on Walmart entry across the benchmark models and throughout both
partial-identification intervals. The South effect for Walmart and the Midwest
effect for Kmart are less precisely identified, with intervals that include
values close to zero. The intercepts imply lower baseline profitability for
Kmart than for Walmart, consistent with the benchmark estimates.

Column (5) accepts a wide range of common unobserved heterogeneity: the retained
common-shock scales imply \(\rho\in[0.00,0.93]\). Thus, the singleton-event
class outer set does not rely on ruling out strong positive correlation in
firms' unobserved profitability shocks. The sampled sharp-set approximation in
Column (6) accepts \(\rho\in[0.00,0.86]\). Comparing Columns (5) and (6), the
core restrictions tighten several intervals, especially for retail sales per
capita, the competitive effect, and the Kmart intercept. The sampled sharp-set
intervals should be interpreted with caution because they are based on accepted
points from a numerical search rather than exact projections; in particular, the
search may miss boundary points and therefore understate some projection ranges.
With this caveat, the comparison suggests that the singleton-event class outer
set recovers many qualitative implications, while the core restrictions provide
additional quantitative tightening. The singleton-event class convex projections
take \(112.3\) seconds of total computational time, while the sampled sharp-set
approximation takes \(14{,}060.9\) seconds, excluding the singleton-class
convex projections used to construct its search region.

%% file: S.08.Empirical.Application.II.tex
\section{Empirical Application II: Burger-Chain Entry Game\label{section:empirical.application.II}}

As a second, more demanding application, I study ordered-entry decisions by McDonald's, Burger King, and Wendy's, the three largest U.S. burger chains. In 2019, these firms accounted for more than 70 percent of sales among the top 20 U.S. burger chains \citep{technomic2019}. I model their decisions as a three-player ordered-action entry game in which each chain chooses how many outlets to operate in a local market. This application extends the analysis beyond binary entry by allowing each firm to choose the intensity of its market presence. Although the model abstracts from fringe competitors, the high sales concentration among the three leading chains makes this a natural setting for studying strategic entry and expansion decisions among major national chains.

\subsection{Setup}

\paragraph{Data Preparation}
My primary data source is the 2019 cross-section of the Data Axle Historical Business Database, which reports outlet locations for McDonald's, Burger King, and Wendy's. Following \citet{koh2023stable}, I define markets as 2010 Census-designated urban tracts.\footnote{Urban census tracts provide a useful local-market unit because quick-service restaurant entry is highly local and because the model studies both entry and outlet intensity. I interpret tracts as empirical markets for modeling local entry and outlet-intensity decisions, not as formal geographic markets for chain-level network expansion.} Each chain's action is the number of outlets it operates in a tract, discretized as $\mathcal{Y}_i=\{0,1,2\}$, where action $Y_i=2$ denotes two or more outlets. Table \ref{table:number.of.outlets.by.firm} reports the distribution of outlet counts by firm.

\begin{table}[htbp!]
    \centering
    \caption{Number of outlets in urban tracts}
    \label{table:number.of.outlets.by.firm}
    \begin{threeparttable}
    \begin{tabular}{lccc}
    \toprule
         Firm $\backslash$ Outlets & $0$ & $1$ & $2^+$ \\ \midrule
         McDonald's & 44,001 & 10,322 & 835 \\
         Burger King & 49,047 & 5,901 & 210 \\
         Wendy's & 50,397 & 4,646 & 115 \\ \bottomrule
    \end{tabular}
    \begin{tablenotes}
        \footnotesize
        \item \emph{Notes}: The table reports the number of urban tracts corresponding to each firm-number-of-outlets pair in 2019.
    \end{tablenotes}
    \end{threeparttable}
\end{table}

To control for market characteristics, I use two covariates. First, I use the number of eating and drinking places in each tract in 2017 from the National Neighborhood Data Archive (NaNDA) \citep{esposito2020nanda}. This variable proxies for local commercial activity and demand conditions relevant for quick-service restaurant entry.\footnote{This control is predetermined relative to the 2019 outlet configuration used in the estimation. I interpret it as capturing broader local market attractiveness, rather than as an outcome determined by the contemporaneous entry decisions of the three burger chains.} Second, I use the number of own-firm outlets per 100,000 county residents as a firm-specific demand shifter, capturing the strength of each chain's local network presence. This county-level network measure is constructed after subtracting the focal tract's own outlets from the county total, so that the outcome in a tract does not mechanically enter its own payoff shifter. Table \ref{table:summary.statistics.burger} reports summary statistics.

\begin{table}[htbp!]
\small
    \centering
    \caption{Summary statistics of burger-chain entry game data}
    \label{table:summary.statistics.burger}
    \begin{threeparttable}
    \begin{tabular}{lccccc}
    \toprule
    Variable & Mean & Median & Std Dev. & Min & Max \\ \midrule
    McDonald's outlets & 0.22 & 0.00 & 0.45 & 0 & 2 \\
    Burger King outlets & 0.11 & 0.00 & 0.33 & 0 & 2 \\
    Wendy's outlets & 0.09 & 0.00 & 0.29 & 0 & 2 \\ 
    Number of eating places & 11.77 & 8.00 & 13.14 & 0 & 311 \\
    MD outlets per 100,000 people in county & 4.42 & 4.46 & 1.44 & 0 & 22.16  \\
    BK outlets per 100,000 people in county & 2.28 & 2.25 & 1.08 & 0 & 17.75 \\
    WD outlets per 100,000 people in county & 1.77 & 1.61 & 1.13 & 0 & 16.87 \\ \midrule
    Number of markets (urban tracts) & 55,158 \\ \bottomrule
    \end{tabular}
    \begin{tablenotes}
        \footnotesize
        \item \emph{Notes}: The unit of observation is a market, defined as a Census-designated urban tract in 2010.
    \end{tablenotes}
    \end{threeparttable}
\end{table}

To construct the covariate support used in the partial-identification exercise, I discretize the number of eating places and the three chain-specific county-network variables into two support points using the median-split procedure described in Section \ref{section:empirical.application}. Each bin is represented by the within-bin mean of the corresponding continuous variable. This yields $|\mathcal{X}|=16$ covariate cells. 

\paragraph{Structural Model}
The structural model is a three-player ordered-entry game. Let $\mathcal{I}=\{\mathrm{MD},\mathrm{BK},\mathrm{WD}\}$ denote the set of chains. Let $E_m$ denote the number of eating and drinking places in market $m$, and let $N_{im}$ denote chain $i$'s leave-one-tract-out county outlet density per 100,000 residents. I specify chain $i$'s payoff as
\begin{equation}\label{equation:chain.entry.game.payoff.function.0}
u_i(y_i,y_{-i},E_m,N_{im},\lambda,\varepsilon_i)
=
y_i
\left(
\delta_i(E_m,N_{im})
-\eta_i y_i
-
\sum_{j\neq i}\Delta_{ij}y_j
+
\sigma\lambda
+
\varepsilon_i
\right),
\end{equation}
where
\begin{equation}\label{equation:chain.entry.game.delta}
\delta_i(E_m,N_{im})
=
\alpha_i
+
\beta_i E_m
+
\rho_i N_{im}.
\end{equation}
The intercept $\alpha_i$, eating-place coefficient $\beta_i$, network coefficient $\rho_i$, and concavity coefficient $\eta_i$ are chain-specific. The competitive effects are directed and pair-specific: $\Delta_{ij}$ measures the effect of an additional outlet of rival chain $j$ on chain $i$'s payoff. I impose $\eta_i\geq 0$ and $\Delta_{ij}\geq 0$, so that payoffs are concave in own outlet counts and rival outlets weakly reduce payoffs. The unobserved payoff component is $\sigma\lambda+\varepsilon_i$, where $\lambda$ is a standard normal market-level shock and $\varepsilon_i$ is a chain-specific standard logistic shock, independent across chains and independent of $\lambda$.

\subsection{Estimation and Inference}

I use the same confidence-set logic as in Section
\ref{section:empirical.application}, but adapt the implementation to the
larger ordered-action outcome space. The event collection is the singleton
class: for each action profile $y\in\mathcal Y=\{0,1,2\}^3$ and covariate cell
$x\in\mathcal X$, I impose $\phi(y\mid x)\leq \mathcal{L}_\theta(y\mid x),$ where $\mathcal L_\theta(y\mid x)$ is the maximum model-implied probability that $y$ is a
pure-strategy Nash equilibrium. Thus, the empirical restrictions are based on
$27$ singleton outcomes in each of the $16$ covariate cells.

The main difference from the first application is the treatment of first-stage
sampling uncertainty. Because many singleton outcome-cell frequencies are small,
I construct one-sided Clopper--Pearson lower confidence bounds for the
singleton probabilities. Let $\underline{\phi}(y\mid x)$ denote the
Bonferroni-adjusted lower bound for $\phi(y\mid x)$. Outcome-cell inequalities
with $\underline{\phi}(y\mid x)=0$ are nonbinding and are omitted from the log
formulation. The resulting outer confidence set is
\[
    \widehat\Theta_I^{1-\alpha}
    =
    \left\{
    \theta:
    \log \underline{\phi}(y\mid x)
    \leq
    \log \mathcal L_\theta(y\mid x),
    \quad
    y\in\mathcal Y,\quad
    x\in\mathcal X,\quad
    \underline{\phi}(y\mid x)>0
    \right\}.
\]
This exact-binomial construction avoids regularizing zero or near-zero log
frequencies and is therefore better suited to the sparse outcome cells in the
ordered-entry application.

The exact singleton generalized likelihood is log-concave in the normalized
payoff parameters conditional on the common-shock scale. In the implementation,
I approximate the common-shock integral by deterministic quadrature and compute
projection intervals from the resulting smooth projection problems over the
retained common-shock values. I evaluate
$\mathcal L_\theta(y\mid x)$ by integrating over the common shock using a 40-point normal
quantile rule, scan over $\sigma\in\{0,0.5,1.0,\ldots,20.0\},$
retain values for which the minimum feasibility slack is below $10^{-5}$, and
compute lower and upper projections for each payoff parameter at the retained
values. The reported intervals are unions over the retained $\sigma$ grid
values. As in the first application, payoff coefficients are reported in the
normalized scale, $\gamma=\theta/\sqrt{\pi^2/3+\sigma^2}.$ Computational details are further described in Appendix \ref{app:computation.burger}.

\subsection{Estimation Results}

Table \ref{table:empirical.application.2.projection.of.confidence.set} reports projection intervals for the singleton-event outer confidence set at the 99\% confidence level. All payoff coefficients are reported on a normalized scale. The bottom rows report the accepted range of implied composite-shock correlations and the total distributed computational time.

\begin{table}[htbp!]
\small
    \centering
    \caption{Empirical results for the burger-chain ordered-entry application}
    \label{table:empirical.application.2.projection.of.confidence.set}
    \begin{threeparttable}
    \begin{tabular}{lccc}
    \toprule
    & \multicolumn{3}{c}{Projection interval ($\alpha = 0.01$)} \\ 
    \cmidrule(lr){2-4}
    Variable & McDonald's & Burger King & Wendy's \\ 
    \midrule
    Intercept ($\alpha_i$) 
        & $[-1.214,-0.989]$ 
        & $[-1.424,-0.960]$ 
        & $[-1.728,-1.239]$ \\
    Eating places ($\beta_i$) 
        & $[0.057,0.062]$ 
        & $[0.052,0.060]$ 
        & $[0.054,0.063]$ \\
    Own outlets per 100,000 county residents ($\rho_i$) 
        & $[-0.029,0.019]$ 
        & $[-0.112,0.000]$ 
        & $[-0.013,0.068]$ \\
    Concavity coefficient ($\eta_i$) 
        & $[0.280,0.433]$ 
        & $[0.135,0.449]$ 
        & $[0.129,0.431]$ \\ 
    \midrule
    \multicolumn{4}{l}{Directed competitive effects, $\Delta_{ij}$} \\
    \; Rival is McDonald's 
        & -- 
        & $[0.279,0.511]$ 
        & $[0.246,0.510]$ \\
    \; Rival is Burger King 
        & $[0.299,0.606]$ 
        & -- 
        & $[0.145,0.507]$ \\
    \; Rival is Wendy's 
        & $[0.227,0.567]$ 
        & $[0.065,0.409]$ 
        & -- \\ 
    \midrule
    Implied shock-correlation range 
        & \multicolumn{3}{c}{$[0.961,0.992]$} \\
    Total runtime (sec.) 
        & \multicolumn{3}{c}{$28{,}227.6$} \\ 
    \bottomrule
    \end{tabular}    
    \begin{tablenotes}
        \footnotesize
        \item \emph{Notes}: Payoff coefficients are normalized by the standard deviation of each player's composite unobserved payoff shock, $\sqrt{\pi^2/3+\sigma^2}$. The directed competitive-effect parameter $\Delta_{ij}$ measures the effect of rival $j$'s outlet count on affected chain $i$'s payoff and enters the payoff with a negative sign. Concavity and competitive-effect coefficients are constrained to be nonnegative. The reported intervals are unions over accepted $\sigma$ grid values. Runtime is the total distributed computational time. 
    \end{tablenotes}
    \end{threeparttable}
\end{table}

The estimates are economically intuitive. The coefficient on eating places is
positive and tightly bounded for all three chains. Ten additional eating places
increase normalized payoffs by roughly \(0.52\) to \(0.63\), depending on the
chain. The intercepts are negative, with McDonald's having the least negative
interval and Wendy's the most negative interval. This pattern is consistent with
stronger baseline profitability for McDonald's at low values of the observed
demand shifters.

The own-network coefficients are less informative. The McDonald's and Wendy's
intervals include zero, while the Burger King interval is weakly negative with
an upper endpoint essentially at zero. Thus, after controlling for local
restaurant activity and allowing for common market-level unobserved
heterogeneity, the singleton-event inequalities do not provide strong evidence
that county-level own-chain networks raise tract-level profitability.

The concavity coefficients are bounded away from zero for all three chains,
implying diminishing returns from operating multiple outlets in the same tract.
The directed competitive effects are also positive across all ordered pairs.
Because these effects enter payoffs with a negative sign, the estimates imply
that rival outlets reduce a chain's payoff. Allowing these effects to be
directed matters conceptually: the effect of Burger King on McDonald's need not
equal the effect of McDonald's on Burger King. In the estimated set, however,
the intervals overlap substantially, so the results support robust adverse
competitive effects without sharply ranking the strength of competition across
directed pairs.

The accepted \(\sigma\) grid indicates that the singleton-event restrictions
require substantial common market-level unobserved heterogeneity. The retained
common-shock scales imply
\(\rho\in[0.961,0.992]\), so the accepted parameter values feature highly
correlated composite payoff shocks across chains.\footnote{The accepted grid
values of the common-shock scale are
\(\sigma\in\{9.0,9.5,\ldots,20.0\}\). Under the normalization
\(\lambda\sim N(0,1)\) and standard-logistic idiosyncratic shocks, the
correlation between two players' composite payoff shocks is
\(\rho(\sigma)=\sigma^2/(\sigma^2+\pi^2/3)\). These accepted grid values
correspond to implied correlations ranging from approximately \(0.961\) to
\(0.992\). Because \(\sigma=20.0\) is the upper endpoint of the scan grid, the
upper endpoint should be interpreted as the accepted range over the grid used in
the computation, not as an estimated upper bound on \(\sigma\).}
This pattern is plausible in this setting because local demand shocks that make
a tract attractive for one burger chain are likely to make it attractive for
rival chains as well.

Computationally, the application illustrates the value of the convex
formulation. The projection problem involves 18 normalized payoff parameters,
and a direct tensor-grid search over these parameters alone would require
\(m^{18}\) membership checks per value of \(\sigma\); even \(m=5\) implies about
\(3.8\times 10^{12}\) checks. The convex approach instead computes the reported
projection intervals through a sequence of tractable optimization problems. The
reported runtime is \(28{,}227.6\) seconds of total distributed computational
time.

%% file: S.99.Conclusion.tex
\section{Conclusion \label{section:conclusion}}

This paper develops a tractable approach to estimating finite
complete-information discrete games without imposing an equilibrium-selection
rule. The central idea is to replace the full collection of sharp
generalized-likelihood inequalities with carefully chosen restrictions that
preserve substantial identifying power while restoring convex structure. Under
affine payoff differences and log-concave payoff shocks, the resulting
singleton-event generalized likelihoods are log-concave in payoff parameters
for each fixed value of the common-shock scale. In unordered- and ordered-action
games, standard logit-type assumptions further deliver closed-form
best-response probabilities. The resulting exact outer sets can be characterized
through fixed-\(\sigma\) convex restrictions, replacing repeated equilibrium
enumeration, simulation over idiosyncratic payoff shocks, and high-dimensional
parameter-space search with tractable projection problems.

The numerical and empirical applications show that the proposed sets can be
informative and computationally scalable. In the Walmart--Kmart application,
the singleton-event outer set delivers informative bounds in \(112.3\) seconds
of total computational time, while a sampled sharp-set approximation on the
reduced search region takes \(3.9\) worker-hours of total distributed
computational time. In the burger-chain application, the ordered-action
procedure delivers informative projection intervals for an 18-dimensional
normalized payoff vector in \(7.8\) worker-hours of total distributed
computational time, corresponding to \(78.9\) minutes of elapsed wall-clock time
on six Julia worker processes. These results suggest that convex
generalized-likelihood restrictions can make moment-inequality methods more
practical in empirical discrete games.

Several directions remain open. First, the logic of the paper may apply beyond
the specific unordered- and ordered-action environments studied here. The
broader lesson is that analytically convenient structures familiar from
single-agent discrete choice can sometimes be recovered in multi-agent
partially identified models by changing the target from the full sharp set to a
carefully chosen tractable implication. Second, further work could improve the
numerical implementation, especially in integrating over common unobserved
heterogeneity and in scaling the method to large network games, as in
\citet{jia_what_2008}. Finally, developing inference methods that better
exploit the convex structure of the proposed restrictions is a natural next
step. The framework developed here provides a computational foundation for such
extensions.

\bigskip
\paragraph{Disclosure Statement}
During the preparation of this work, the author used OpenAI's ChatGPT 5.5 to
improve the clarity of the writing and to assist with coding and debugging.
After using this tool, the author reviewed and edited the content as needed and
takes full responsibility for the published article. Wharton Research Data
Services (WRDS) was used to prepare part of the data set reported in this
manuscript. This service and the data available thereon constitute valuable
intellectual property and trade secrets of WRDS and/or its third-party
suppliers. The author declares that no financial or personal conflicts of
interest influenced the research presented in this paper.

\paragraph{Data Availability Statement} 
Replication code and non-proprietary replication materials are available from the author upon request. The Walmart--Kmart application uses the public data from \citet{jia_what_2008} and \citet{ellickson_structural_2011}. The burger-chain application uses the Data Axle Historical Business Database, which is proprietary and was accessed through Wharton Research Data Services (WRDS). Restrictions apply to the availability of the Data Axle data, which were used under license and cannot be redistributed by the author. Researchers with appropriate WRDS/Data Axle access can use the replication code to reconstruct the analysis dataset. The remaining public covariates used in the burger-chain application are available from the public sources cited in the article.

%% file: S.A.Proofs.tex
\section{Proofs \label{section:proofs}}

\subsection{Proof of Theorem \ref{theorem:primitive.log.concavity.singleton}}
\label{section:proof.of.theorem.2}

Fix $\sigma\in\Sigma$, $y\in\mathcal Y$, and $x\in\mathcal X$. Throughout the proof, log-concavity of nonnegative functions is understood in the extended-value sense, so that zeros are allowed and the indicator of a convex set is log-concave. I first show
that each player-level best-response probability is log-concave in
$(\gamma,\lambda)$.

For each player $i$, define the set of parameter--shock values under which
the prescribed action $y_i$ is a best response to $y_{-i}$:
\[
B_i^\sigma(y,x)
\equiv
\left\{
(\gamma,\lambda,\varepsilon_i):
D_{i,a_i}^{(\gamma,\sigma)}(y,x,\lambda,\varepsilon_i)\geq 0
\text{ for all } a_i\in\mathcal Y_i
\right\}.
\]
The inequality for $a_i=y_i$ is redundant, since the corresponding payoff
difference is zero, but including it is harmless.

By Assumption \ref{assumption:additive.separability.linear.payoffs}, for each
$a_i\in\mathcal Y_i$,
\[
D_{i,a_i}^{(\gamma,\sigma)}(y,x,\lambda,\varepsilon_i)
=
v_i^\gamma(y_i,y_{-i},x)
-
v_i^\gamma(a_i,y_{-i},x) 
+
\left[
r_i(y_i)-r_i(a_i)
\right]^\top S_i(\sigma)\lambda
+
\left[
r_i(y_i)-r_i(a_i)
\right]^\top \varepsilon_i .
\]
For fixed $\sigma$, this expression is affine in
$(\gamma,\lambda,\varepsilon_i)$. Hence each no-deviation inequality defines
a halfspace in $(\gamma,\lambda,\varepsilon_i)$, and
$B_i^\sigma(y,x)$ is a finite intersection of halfspaces. Therefore
$B_i^\sigma(y,x)$ is convex.

Now fix $(\gamma,\lambda)$ and consider the section of this set in the
idiosyncratic shock dimension:
\[
B_i^\sigma(y,x;\gamma,\lambda)
\equiv
\left\{
\varepsilon_i:
(\gamma,\lambda,\varepsilon_i)\in B_i^\sigma(y,x)
\right\}.
\]
This section is the set of player $i$'s idiosyncratic shocks under which
$y_i$ is a best response to $y_{-i}$. Therefore the player-level
best-response probability can be written as
\[
\begin{split}
\mathcal M_i^{(\gamma,\sigma)}(y_i\mid y_{-i},x,\lambda)
& =
\int
\mathbb I\{
\varepsilon_i\in B_i^\sigma(y,x;\gamma,\lambda)
\}
q_i(\varepsilon_i)d\varepsilon_i \\
& =
\int
\mathbb I\{(\gamma,\lambda,\varepsilon_i)\in B_i^\sigma(y,x)\}
q_i(\varepsilon_i)d\varepsilon_i .
\end{split}
\]

Since \(B_i^\sigma(y,x)\) is convex, the indicator $\mathbb I\{(\gamma,\lambda,\varepsilon_i)\in B_i^\sigma(y,x)\}$ is log-concave in the extended-value sense. Since \(q_i(\varepsilon_i)\) is log-concave in
\(\varepsilon_i\), and can be viewed as a function of \((\gamma,\lambda,\varepsilon_i)\) that is
constant in \((\gamma,\lambda)\), the function $H_i(\gamma,\lambda,\varepsilon_i)
    \equiv
    \mathbb I\{(\gamma,\lambda,\varepsilon_i)\in B_i^\sigma(y,x)\}
    q_i(\varepsilon_i)$
is log-concave in \((\gamma,\lambda,\varepsilon_i)\). By Prékopa's theorem, marginalizing a
log-concave function preserves log-concavity. Hence
\[
    M_i^{(\gamma,\sigma)}(y_i\mid y_{-i},x,\lambda)
    =
    \int H_i(\gamma,\lambda,\varepsilon_i)\,d\varepsilon_i
\]
is log-concave in \((\gamma,\lambda)\).

The player-level sections also make the equilibrium event transparent. For fixed
\((\gamma,\lambda)\), the profile \(y\) is a pure-strategy Nash equilibrium if and only if every
player passes her best-response test:
\[
    \{\varepsilon : y\in G^{(\gamma,\sigma)}(x,\lambda,\varepsilon)\}
    =
    \bigcap_{i=1}^I
    \{\varepsilon : \varepsilon_i\in B_i^\sigma(y,x;\gamma,\lambda)\}
    =
    \mathop{\times}_{i=1}^I B_i^\sigma(y,x;\gamma,\lambda).
\]
The last equality is a Cartesian product: conditional on \((x,\lambda)\) and the candidate profile
\(y\), player \(i\)'s best-response test depends on \(\varepsilon_i\), but not on
\(\varepsilon_j\) for \(j\neq i\). By Assumption~2, the idiosyncratic shocks are independent of
\((x,\lambda)\) and mutually independent across players. Therefore,
\[
\begin{aligned}
    \mathcal L_{(\gamma,\sigma)}(y\mid x,\lambda)
    &=
    \Pr\left(
        \varepsilon\in \mathop{\times}_{i=1}^I B_i^\sigma(y,x;\gamma,\lambda)
        \mid x,\lambda
    \right) \\
    &=
    \prod_{i=1}^I
    \Pr\left(
        \varepsilon_i\in B_i^\sigma(y,x;\gamma,\lambda)
        \mid x,\lambda
    \right) \\
    &=
    \prod_{i=1}^I
    M_i^{(\gamma,\sigma)}(y_i\mid y_{-i},x,\lambda).
\end{aligned}
\]
Because products of nonnegative log-concave functions are log-concave,
\(\mathcal L_{(\gamma,\sigma)}(y\mid x,\lambda)\) is log-concave in \((\gamma,\lambda)\).

It remains to integrate out the common unobservable \(\lambda\). By the law of iterated
expectations,
\[
    \mathcal L_{(\gamma,\sigma)}(y\mid x)
    =
    \int \mathcal L_{(\gamma,\sigma)}(y\mid x,\lambda)h(\lambda)\,d\lambda .
\]
The conditional likelihood \(\mathcal L_{(\gamma,\sigma)}(y\mid x,\lambda)\) is log-concave in
\((\gamma,\lambda)\). By Assumption~2, \(h(\lambda)\) is log-concave in \(\lambda\), and it can be
viewed as a function of \((\gamma,\lambda)\) that is constant in \(\gamma\). Hence $\mathcal L_{(\gamma,\sigma)}(y\mid x,\lambda)h(\lambda)$
is log-concave in \((\gamma,\lambda)\). Applying Prékopa's theorem again, now marginalizing over
\(\lambda\), implies that \(\mathcal L_{(\gamma,\sigma)}(y\mid x)\) is log-concave in \(\gamma\).
\qed

\subsection{Proof of Theorem~\ref{theorem:unordered.singleton.outer.set}}
\label{section:proof.of.theorem.3}

Fix $\sigma\in\Sigma$, $y\in\mathcal Y$, $x\in\mathcal X$, and $\lambda$.
For player $i$, conditional on $(x,\lambda)$ and on opponents' actions
$y_{-i}$, the payoff from action $a_i\in\mathcal Y_i$ is $u_i^{(\gamma,\sigma)}(a_i,y_{-i},x,\lambda,\varepsilon_i)
=
V_i^{(\gamma,\sigma)}(a_i,y_{-i},x,\lambda)
+
\varepsilon_i(a_i).$
Therefore,
\[
\begin{split}
\mathcal M_i^{(\gamma,\sigma)}(y_i\mid y_{-i},x,\lambda)
=
\Pr\Big(
&
V_i^{(\gamma,\sigma)}(y_i,y_{-i},x,\lambda)
+
\varepsilon_i(y_i) \\
&\geq
V_i^{(\gamma,\sigma)}(a_i,y_{-i},x,\lambda)
+
\varepsilon_i(a_i)
\quad
\text{for all } a_i\in\mathcal Y_i
\mid x,\lambda
\Big).
\end{split}
\]
Under Assumption~\ref{assumption:type.1.extreme.value.unordered}, the
action-specific payoff shocks are independent and identically distributed
Type-I extreme-value random variables. The standard multinomial-logit formula
then gives
\[
\mathcal M_i^{(\gamma,\sigma)}(y_i\mid y_{-i},x,\lambda)
=
\frac{
\exp\left(
V_i^{(\gamma,\sigma)}(y_i,y_{-i},x,\lambda)
\right)
}{
\sum_{a_i\in\mathcal Y_i}
\exp\left(
V_i^{(\gamma,\sigma)}(a_i,y_{-i},x,\lambda)
\right)
}.
\]
Because the Type-I extreme-value distribution is continuous, ties occur with
probability zero, so weak and strict best-response inequalities yield the
same probability.

Next, by the definition of pure-strategy Nash equilibrium, $y$ is an
equilibrium if and only if every player $i$'s prescribed action $y_i$ is a
best response to $y_{-i}$. Conditional on $(x,\lambda)$, the idiosyncratic
shock vectors $\varepsilon_1,\ldots,\varepsilon_I$ are mutually independent.
Hence,
\[
\mathcal L_{(\gamma,\sigma)}(y\mid x,\lambda)
=
\prod_{i=1}^I
\mathcal M_i^{(\gamma,\sigma)}(y_i\mid y_{-i},x,\lambda).
\]
Substituting the multinomial-logit expression for each player-level
best-response probability yields
\[
\mathcal L_{(\gamma,\sigma)}(y\mid x,\lambda)
=
\prod_{i=1}^I
\frac{
\exp\left(
V_i^{(\gamma,\sigma)}(y_i,y_{-i},x,\lambda)
\right)
}{
\sum_{a_i\in\mathcal Y_i}
\exp\left(
V_i^{(\gamma,\sigma)}(a_i,y_{-i},x,\lambda)
\right)
}.
\]

The unconditional generalized likelihood follows by integrating out the
common shock. Since $\lambda$ is independent of $x$ and has density $h$,
\[
\mathcal L_{(\gamma,\sigma)}(y\mid x)
=
\int
\mathcal L_{(\gamma,\sigma)}(y\mid x,\lambda)h(\lambda)\,d\lambda .
\]
Using the previous display gives
\[
\mathcal L_{(\gamma,\sigma)}(y\mid x)
=
\int
\prod_{i=1}^I
\frac{
\exp\left(
V_i^{(\gamma,\sigma)}(y_i,y_{-i},x,\lambda)
\right)
}{
\sum_{a_i\in\mathcal Y_i}
\exp\left(
V_i^{(\gamma,\sigma)}(a_i,y_{-i},x,\lambda)
\right)
}
h(\lambda)\,d\lambda .
\]

The remaining claims follow directly from
Theorem~\ref{theorem:primitive.log.concavity.singleton}. The assumptions of
that theorem are maintained here, and
Assumption~\ref{assumption:type.1.extreme.value.unordered} only adds the
Type-I extreme-value structure needed for the closed-form multinomial-logit
expression above. Hence, for each fixed \(\sigma\),
\(\mathcal M_i^{(\gamma,\sigma)}(y_i\mid y_{-i},x,\lambda)\) and
\(\mathcal L_{(\gamma,\sigma)}(y\mid x,\lambda)\) are log-concave in
\((\gamma,\lambda)\), and
\(\mathcal L_{(\gamma,\sigma)}(y\mid x)\) is log-concave in \(\gamma\).

Finally, suppose \(\Gamma\) is convex. For fixed \(\sigma\), the slice of the
singleton-event outer set is
\[
\Gamma_I^*(\sigma)
=
\Gamma
\cap
\bigcap_{(y,x)\in\mathcal Y\times\mathcal X}
\left\{
\gamma:
\phi(y\mid x)
\leq
\mathcal L_{(\gamma,\sigma)}(y\mid x)
\right\}.
\]
If \(\phi(y\mid x)=0\), the corresponding restriction is vacuous. If
\(\phi(y\mid x)>0\), the restriction is a superlevel-set restriction for the
log-concave function \(\mathcal L_{(\gamma,\sigma)}(y\mid x)\). Since
superlevel sets of log-concave functions are convex, each nonvacuous
restriction defines a convex feasible set in \(\gamma\). Intersecting these
sets with the convex parameter space \(\Gamma\) preserves convexity. Therefore
\(\Gamma_I^*(\sigma)\) is convex.

\subsection{Proof of Lemma~\ref{lemma:threshold.crossing.rule}}
\label{section:proof.of.lemma.threshold.crossing.rule}

Fix player $i$, opponents' action profile $y_{-i}$, covariates $x$, common
shock $\lambda$, and parameters $(\gamma,\sigma)$. For notational simplicity,
write
\[
V_\ell
\equiv
V_i^{(\gamma,\sigma)}(s_i^\ell,y_{-i},x,\lambda),
\qquad
d_\ell(\varepsilon_i)
\equiv
\frac{
u_i^{(\gamma,\sigma)}(s_i^\ell,y_{-i},x,\lambda,\varepsilon_i)
-
u_i^{(\gamma,\sigma)}(s_i^{\ell-1},y_{-i},x,\lambda,\varepsilon_i)
}{
s_i^\ell-s_i^{\ell-1}
}
\]
for $\ell=1,\ldots,L_i$. Since $u_i^{(\gamma,\sigma)}(s_i^\ell,y_{-i},x,\lambda,\varepsilon_i)
=
V_\ell+s_i^\ell\varepsilon_i,$
we have
\[
d_\ell(\varepsilon_i)
=
\frac{V_\ell-V_{\ell-1}}{s_i^\ell-s_i^{\ell-1}}
+\varepsilon_i
=
\varepsilon_i
-
e_i^{(\gamma,\sigma)}(s_i^\ell,y_{-i},x,\lambda).
\]
Strict concavity of $V_i^{(\gamma,\sigma)}(\cdot,y_{-i},x,\lambda)$ on the
ordered action grid implies that the adjacent slopes
\[
\frac{V_\ell-V_{\ell-1}}{s_i^\ell-s_i^{\ell-1}},
\qquad \ell=1,\ldots,L_i,
\]
are strictly decreasing in $\ell$. Therefore the thresholds, which are the
negatives of these adjacent slopes, are strictly increasing:
\[
e_i^{(\gamma,\sigma)}(s_i^1,y_{-i},x,\lambda)
<
\cdots
<
e_i^{(\gamma,\sigma)}(s_i^{L_i},y_{-i},x,\lambda).
\]
Together with the boundary definitions
$e_i^{(\gamma,\sigma)}(s_i^0,y_{-i},x,\lambda)=-\infty$ and
$e_i^{(\gamma,\sigma)}(s_i^{L_i+1},y_{-i},x,\lambda)=+\infty$, this proves
the stated ordering of thresholds.

Now fix $\varepsilon_i$ and consider action $s_i^\ell$. The action
$s_i^\ell$ weakly dominates the adjacent lower action $s_i^{\ell-1}$ if and
only if $d_\ell(\varepsilon_i)\geq 0$, equivalently $\varepsilon_i
\geq
e_i^{(\gamma,\sigma)}(s_i^\ell,y_{-i},x,\lambda).$
Similarly, it weakly dominates the adjacent higher action $s_i^{\ell+1}$ if
and only if $d_{\ell+1}(\varepsilon_i)\leq 0$, equivalently $\varepsilon_i
\leq
e_i^{(\gamma,\sigma)}(s_i^{\ell+1},y_{-i},x,\lambda).$
For boundary actions, the lower comparison is omitted when $\ell=0$ and the
upper comparison is omitted when $\ell=L_i$; equivalently, these cases are
covered by the conventions
$e_i^{(\gamma,\sigma)}(s_i^0,y_{-i},x,\lambda)=-\infty$ and
$e_i^{(\gamma,\sigma)}(s_i^{L_i+1},y_{-i},x,\lambda)=+\infty$.

It remains only to verify that adjacent comparisons characterize global
optimality. First consider an interior action $s_i^\ell$, where
$1\leq \ell\leq L_i-1$. If $e_i^{(\gamma,\sigma)}(s_i^\ell,y_{-i},x,\lambda)
\leq
\varepsilon_i
\leq
e_i^{(\gamma,\sigma)}(s_i^{\ell+1},y_{-i},x,\lambda),$
then $d_\ell(\varepsilon_i)\geq 0$ and
$d_{\ell+1}(\varepsilon_i)\leq 0$. Since strict concavity implies that
$d_m(\varepsilon_i)$ is strictly decreasing in $m$, it follows that
$d_m(\varepsilon_i)\geq 0$ for all $m\leq \ell$ and
$d_m(\varepsilon_i)\leq 0$ for all $m\geq \ell+1$. Hence payoffs weakly
increase as the action moves upward toward $s_i^\ell$ and weakly decrease as
the action moves above $s_i^\ell$. Thus $s_i^\ell$ is a best response to
$y_{-i}$.

The boundary cases are analogous. If $\ell=0$, the condition $\varepsilon_i
\leq
e_i^{(\gamma,\sigma)}(s_i^1,y_{-i},x,\lambda)$
implies $d_m(\varepsilon_i)\leq 0$ for all $m\geq 1$, so payoffs weakly
decrease as the action rises above $s_i^0$. Hence $s_i^0$ is a best response.
If $\ell=L_i$, the condition $\varepsilon_i
\geq
e_i^{(\gamma,\sigma)}(s_i^{L_i},y_{-i},x,\lambda)$
implies $d_m(\varepsilon_i)\geq 0$ for all $m\leq L_i$, so payoffs weakly
increase up to $s_i^{L_i}$. Hence $s_i^{L_i}$ is a best response.

Conversely, if $s_i^\ell$ is a best response, it must weakly dominate each
adjacent action that exists. Therefore, when $\ell\geq 1$,
$d_\ell(\varepsilon_i)\geq 0$, which is equivalent to $\varepsilon_i
\geq
e_i^{(\gamma,\sigma)}(s_i^\ell,y_{-i},x,\lambda),$
and when $\ell\leq L_i-1$, $d_{\ell+1}(\varepsilon_i)\leq 0$, which is
equivalent to $\varepsilon_i
\leq
e_i^{(\gamma,\sigma)}(s_i^{\ell+1},y_{-i},x,\lambda).$
The boundary conventions
$e_i^{(\gamma,\sigma)}(s_i^0,y_{-i},x,\lambda)=-\infty$ and
$e_i^{(\gamma,\sigma)}(s_i^{L_i+1},y_{-i},x,\lambda)=+\infty$ make these
conditions valid for all $\ell=0,\ldots,L_i$.

Therefore, for any action profile $y$, each $y_i$ is a best response to
$y_{-i}$ if and only if $e_i^{(\gamma,\sigma)}(y_i,y_{-i},x,\lambda)
\leq
\varepsilon_i
\leq
e_i^{(\gamma,\sigma)}(y_i^+,y_{-i},x,\lambda)$
for every player $i$. This is exactly the condition that $y$ is a
pure-strategy Nash equilibrium. If the relevant finite inequalities are
strict, then every lower or higher action yields strictly lower payoff, so
$y_i$ is player $i$'s unique best response to $y_{-i}$. \qed

\subsection{Proof of Theorem~\ref{theorem:ordered.singleton.outer.set}}
\label{section:proof.of.theorem.ordered.singleton.outer.set}

Fix $\sigma\in\Sigma$, an action profile $y$, covariate value $x$, common shock
$\lambda$, and player $i$. By Lemma~\ref{lemma:threshold.crossing.rule}, conditional on
$(x,\lambda)$ and opponents' actions $y_{-i}$, action $y_i$ is a best response
if and only if
\[
e_i^{(\gamma,\sigma)}(y_i,y_{-i},x,\lambda)
\leq
\varepsilon_i
\leq
e_i^{(\gamma,\sigma)}(y_i^+,y_{-i},x,\lambda).
\]
Under Assumption~\ref{assumption:logistic.distribution.ordered}, with the
conventions $F(-\infty)=0$ and $F(+\infty)=1$, this gives
\[
\mathcal M_i^{(\gamma,\sigma)}(y_i\mid y_{-i},x,\lambda)
=
F\!\left(
e_i^{(\gamma,\sigma)}(y_i^+,y_{-i},x,\lambda)
\right)
-
F\!\left(
e_i^{(\gamma,\sigma)}(y_i,y_{-i},x,\lambda)
\right).
\]
Because the idiosyncratic shocks are independent across players conditional
on $(x,\lambda)$, the conditional singleton-event generalized likelihood is
the product of the player-level best-response probabilities:
\[
\mathcal L_{(\gamma,\sigma)}(y\mid x,\lambda)
=
\prod_{i=1}^I
\left[
F\!\left(
e_i^{(\gamma,\sigma)}(y_i^+,y_{-i},x,\lambda)
\right)
-
F\!\left(
e_i^{(\gamma,\sigma)}(y_i,y_{-i},x,\lambda)
\right)
\right].
\]
Integrating over the common shock gives
\[
\mathcal L_{(\gamma,\sigma)}(y\mid x)
=
\int
\prod_{i=1}^I
\left[
F\!\left(
e_i^{(\gamma,\sigma)}(y_i^+,y_{-i},x,\lambda)
\right)
-
F\!\left(
e_i^{(\gamma,\sigma)}(y_i,y_{-i},x,\lambda)
\right)
\right]
h(\lambda)\,d\lambda .
\]

The remaining log-concavity claim follows directly from
Theorem~\ref{theorem:primitive.log.concavity.singleton}. The assumptions of
that theorem are maintained here. Assumption~\ref{assumption:strict.concavity}
and Lemma~\ref{lemma:threshold.crossing.rule} provide the threshold
representation of best responses, while
Assumption~\ref{assumption:logistic.distribution.ordered} gives the
closed-form ordered-logit interval probability above. Therefore, for each
fixed $\sigma$,
$\mathcal M_i^{(\gamma,\sigma)}(y_i\mid y_{-i},x,\lambda)$ and
$\mathcal L_{(\gamma,\sigma)}(y\mid x,\lambda)$ are log-concave in
$(\gamma,\lambda)$, and
$\mathcal L_{(\gamma,\sigma)}(y\mid x)$ is log-concave in $\gamma$. The
convexity of $\Gamma_I^*(\sigma)$ when $\Gamma$ is convex then follows from
the same superlevel-set argument used for
Theorem~\ref{theorem:primitive.log.concavity.singleton}. \qed

%% file: S.B.Extension.to.Games.with.Vector.Decisions.tex
\section{Extension to Games with Vector Decisions}
\label{section:extension.to.vector.decisions}

This appendix extends the singleton-event construction to games in which each
player chooses a vector of discrete actions, as in \citet{fan2025estimating}.
The coordinates may be unordered choices, such as product-entry indicators, or
ordered choices, such as capacity or store-count decisions. Suppose player $i$
chooses
\[
y_i=(y_{ij})_{j\in\mathcal J_i},
\qquad
\mathcal Y_i=\times_{j\in\mathcal J_i}\mathcal Y_{ij}.
\]
Because $\mathcal Y_i$ is finite, one can always treat the entire vector
$y_i$ as a single action and apply the primitive results in Section
\ref{section:convex.outer.set} directly. The practical difficulty is that the
number of vector actions grows quickly with the number of coordinates. I
therefore describe a coordinate-deviation construction that provides a
tractable outer-set approximation. The construction coincides with the
singleton-event generalized likelihood when coordinatewise optimality
characterizes full best responses.

For a candidate profile $y$, define a coordinate deviation by player $i$ in
coordinate $j$ as
\[
(a_{ij},y_{i,-j},y_{-i}),
\qquad
a_{ij}\in\mathcal Y_{ij},
\]
where $y_{i,-j}$ denotes player $i$'s remaining coordinates. If $y$ is a
pure-strategy Nash equilibrium, then no player can profitably change any
single coordinate:
\[
u_i^\theta(y,x,\lambda,\varepsilon_i)
\geq
u_i^\theta((a_{ij},y_{i,-j}),y_{-i},x,\lambda,\varepsilon_i),
\quad
\forall i,\ j\in\mathcal J_i,\ a_{ij}\in\mathcal Y_{ij}.
\]
These coordinate-deviation restrictions are necessary for equilibrium.\footnote{They are sufficient only under additional structure ensuring that
single-coordinate local optimality is globally determining. A simple sufficient
condition is additive separability of player \(i\)'s payoff across own action
coordinates, conditional on opponents' actions and covariates. More generally,
one needs a discrete-concavity condition under which absence of profitable
single-coordinate deviations implies global optimality over \(\mathcal Y_i\).}

For unordered coordinates, a closed-form coordinate-deviation probability is
available under coordinate-action-specific Type-I extreme-value shocks. Suppose
player $i$'s payoff can be written as
\[
u_i^\theta(y,x,\lambda,\varepsilon_i)
=
V_i^\theta(y,x,\lambda)
+
\sum_{j\in\mathcal J_i}\varepsilon_{ij}(y_{ij}),
\]
where, conditional on $(x,\lambda)$, the shocks
$\{\varepsilon_{ij}(a): i=1,\ldots,I,\ j\in\mathcal J_i,\ a\in\mathcal Y_{ij}\}$
are mutually independent Type-I extreme-value random variables. For coordinate
$j\in\mathcal J_i$, define
\[
\mathcal P_{ij}^{\theta}(y_{ij}\mid y_{i,-j},y_{-i},x,\lambda)
\equiv
\Pr\left(
u_i^\theta(y,x,\lambda,\varepsilon_i)
\geq
u_i^\theta((a_{ij},y_{i,-j}),y_{-i},x,\lambda,\varepsilon_i)
\quad
\forall a_{ij}\in\mathcal Y_{ij}
\mid x,\lambda
\right).
\]
In this comparison, all coordinate-level shocks except those associated with
coordinate $j$ cancel. Therefore the standard multinomial-logit formula gives
\[
\mathcal P_{ij}^{\theta}(y_{ij}\mid y_{i,-j},y_{-i},x,\lambda)
=
\frac{
\exp\left(V_i^\theta(y,x,\lambda)\right)
}{
\sum_{a_{ij}\in\mathcal Y_{ij}}
\exp\left(
V_i^\theta((a_{ij},y_{i,-j}),y_{-i},x,\lambda)
\right)
}.
\]

For ordered coordinates, the same coordinate-deviation construction applies
with the ordered-logit probability from Section~\ref{section:ordered.actions}
replacing the multinomial-logit probability above. Holding
$(y_{i,-j},y_{-i},x,\lambda)$ fixed, coordinate $j$ is treated as a one-dimensional
ordered-action problem. Under the scalar-shock ordered-action specification and
the corresponding coordinate-level strict concavity condition, the event that
the observed coordinate action is optimal is a threshold interval for the
coordinate shock. With a logistic scalar shock, the resulting coordinate
probability is the ordered-logit interval probability from
Theorem~\ref{theorem:ordered.singleton.outer.set}. Thus the notation
\(\mathcal P_{ij}^{\theta}\) below can denote either the unordered multinomial-logit
coordinate probability or its ordered-logit analogue.

The event that player $i$ passes all coordinate-deviation tests is the
intersection of these coordinate events. Under either the unordered
coordinate-action shock specification or the ordered scalar-shock specification,
the coordinate event for $j$ depends only on the shock or shocks associated with
coordinate $j$. Since these coordinate-level shocks are mutually independent
conditional on $(x,\lambda)$, the coordinate-deviation probabilities multiply:
\[
\overline{\mathcal M}_{i}^{\mathrm{cd},\theta}
(y_i\mid y_{-i},x,\lambda)
=
\prod_{j\in\mathcal J_i}
\mathcal P_{ij}^{\theta}(y_{ij}\mid y_{i,-j},y_{-i},x,\lambda).
\]
Likewise, conditional independence across players gives the conditional
coordinate-deviation upper bound
\[
\overline{\mathcal L}_{\theta}^{\mathrm{cd}}(y\mid x,\lambda)
=
\prod_{i=1}^I
\overline{\mathcal M}_{i}^{\mathrm{cd},\theta}
(y_i\mid y_{-i},x,\lambda).
\]

This object is an upper bound on the singleton-event generalized likelihood.
Indeed, if $y$ is a pure-strategy Nash equilibrium, then each player is a full
best response to $y_{-i}$, and hence each player also passes all coordinate
no-deviation tests. Therefore,
\[
\left\{
y\in G_\theta(x,\lambda,\varepsilon)
\right\}
\subseteq
\left\{
\text{all coordinate-deviation inequalities hold at }y
\right\},
\]
which implies
\[
\mathcal L_{\theta}(y\mid x,\lambda)
\leq
\overline{\mathcal L}_{\theta}^{\mathrm{cd}}(y\mid x,\lambda).
\]
After integrating out the common shock, define
\[
\overline{\mathcal L}_{\theta}^{\mathrm{cd}}(y\mid x)
\equiv
\int
\overline{\mathcal L}_{\theta}^{\mathrm{cd}}(y\mid x,\lambda)
h(\lambda)d\lambda.
\]
Then the inequalities
\[
\phi(y\mid x)
\leq
\overline{\mathcal L}_{\theta}^{\mathrm{cd}}(y\mid x),
\qquad
\forall y\in\mathcal Y,\ x\in\mathcal X,
\]
define a valid outer set.\footnote{If coordinatewise optimality characterizes full
best responses, the inclusion above becomes equality, and $\mathcal L_{\theta}(y\mid x,\lambda)
=
\overline{\mathcal L}_{\theta}^{\mathrm{cd}}(y\mid x,\lambda).$} 

The same convexity logic applies. For unordered coordinates, if for each fixed
\(\sigma\) the functions
\[
V_i^{(\gamma,\sigma)}((a_{ij},y_{i,-j}),y_{-i},x,\lambda)
\]
are affine in \((\gamma,\lambda)\), then each multinomial-logit coordinate
probability is log-concave in \((\gamma,\lambda)\), by the same log-sum-exp
argument used for Theorem~\ref{theorem:unordered.singleton.outer.set}. For
ordered coordinates, the analogous conclusion follows from the threshold
representation and the primitive Pr\'ekopa argument used for
Theorem~\ref{theorem:ordered.singleton.outer.set}. Products preserve
log-concavity, so
\(\overline{\mathcal L}_{(\gamma,\sigma)}^{\mathrm{cd}}(y\mid x,\lambda)\)
is log-concave in \((\gamma,\lambda)\). Multiplying by the log-concave density
\(h(\lambda)\) and applying Pr\'ekopa's theorem implies that
\[
\overline{\mathcal L}_{(\gamma,\sigma)}^{\mathrm{cd}}(y\mid x)
=
\int
\overline{\mathcal L}_{(\gamma,\sigma)}^{\mathrm{cd}}(y\mid x,\lambda)
h(\lambda)d\lambda
\]
is log-concave in \(\gamma\). Therefore, if \(\Gamma\) is convex, each fixed-\(\sigma\)
coordinate-deviation slice
\[
\overline\Gamma_I^{\mathrm{cd}}(\sigma)
=
\left\{
\gamma\in\Gamma:
\phi(y\mid x)
\leq
\overline{\mathcal L}_{(\gamma,\sigma)}^{\mathrm{cd}}(y\mid x),
\quad
\forall y\in\mathcal Y,\ x\in\mathcal X
\right\}
\]
is convex, because it is the intersection of \(\Gamma\) with superlevel sets of
log-concave functions.

This construction is useful in product-entry environments related to
\citet{fan2025estimating}, where firms choose vectors of binary product-entry
decisions, and in ordered vector-decision environments, where each coordinate
may represent an ordered choice such as capacity, quality, or outlet count.
In the binary product-entry case, the coordinate-deviation probability compares
the profitability of the observed product decision to adding or dropping a
single product while holding the firm's other product decisions and rivals'
actions fixed. In the ordered-coordinate case, it compares the observed
coordinate action to alternative ordered values of that coordinate, again
holding the remaining coordinates and rivals' actions fixed. Without assumptions
ensuring that coordinate deviations characterize full best responses, the
resulting inequalities should be interpreted as tractable outer-set restrictions
rather than as the sharp singleton-event generalized likelihood.

%% file: S.C.Additional.Illustrations.for.Ordered.Actions.tex
\section{Additional Illustrations for Ordered Actions}
\label{appendix:ordered.action.illustrations}

This appendix provides additional illustrations for the ordered-action
threshold characterization in Section \ref{section:ordered.actions}. The
examples show how the threshold representation translates equilibrium checks
into interval restrictions on scalar payoff shocks, and how these intervals
lead to closed-form singleton-event generalized likelihoods under logistic
shocks.

\subsection{Thresholds in the Binary-Action Example}

I first illustrate the threshold representation in the running two-player
binary-entry game. Throughout this appendix, I write
$\theta=(\gamma,\sigma)$ when no distinction between the components of the
parameter is needed. Let $\mathcal Y_i=\{0,1\}$ and consider the payoff
specification
\[
u_i^\theta(y_i,y_j,x,\lambda,\varepsilon_i)
=
y_i\left(x_i^\top\beta_i+\Delta_i y_j+\sigma\lambda+\varepsilon_i\right).
\]
The unique finite threshold for player $i$ is
\[
e_i^\theta(1,y_j,x,\lambda)
=
-x_i^\top\beta_i-\Delta_i y_j-\sigma\lambda.
\]
For convenience, define
\[
\underline{\epsilon}_i^\theta(x,\lambda)
\equiv
-x_i^\top\beta_i-\sigma\lambda,
\qquad
\overline{\epsilon}_i^\theta(x,\lambda)
\equiv
\underline{\epsilon}_i^\theta(x,\lambda)-\Delta_i .
\]
Thus, when the opponent does not enter, the cutoff is
$\underline{\epsilon}_i^\theta(x,\lambda)$; when the opponent enters, the
cutoff is $\overline{\epsilon}_i^\theta(x,\lambda)$. If $\Delta_i<0$, the
cutoff is higher when the opponent enters, reflecting strategic substitutes.

Table \ref{table:binary.action.thresholds} summarizes the threshold
representation. The boundary thresholds are
$e_i^\theta(0,y_j,x,\lambda)=-\infty$ and
$e_i^\theta(2,y_j,x,\lambda)=+\infty$. Therefore $y_i=0$ is a best response when
$\varepsilon_i$ lies below the finite cutoff, while $y_i=1$ is a best
response when $\varepsilon_i$ lies above it.

\begin{table}[htbp!]
\centering
\caption{Thresholds and best-response regions in the binary-action example}
\label{table:binary.action.thresholds}
\begin{threeparttable}
\begin{tabular}{cccc}
\toprule
Opponent action $y_j$
& $e_i^\theta(0,y_j,x,\lambda)$
& $e_i^\theta(1,y_j,x,\lambda)$
& $e_i^\theta(2,y_j,x,\lambda)$ \\
\midrule
$0$
& $-\infty$
& $\underline{\epsilon}_i^\theta(x,\lambda)$
& $+\infty$ \\
$1$
& $-\infty$
& $\overline{\epsilon}_i^\theta(x,\lambda)$
& $+\infty$ \\
\bottomrule
\end{tabular}
\begin{tablenotes}
\footnotesize
\item \emph{Notes}: For a given opponent action $y_j$, player $i$ chooses
$0$ if
$e_i^\theta(0,y_j,x,\lambda)\leq \varepsilon_i\leq e_i^\theta(1,y_j,x,\lambda)$
and chooses $1$ if
$e_i^\theta(1,y_j,x,\lambda)\leq \varepsilon_i\leq e_i^\theta(2,y_j,x,\lambda)$.
With continuous shocks, the treatment of threshold endpoints is irrelevant
for probabilities.
\end{tablenotes}
\end{threeparttable}
\end{table}

The table also illustrates how the equilibrium regions in Figure
\ref{PSNE fig:structure} arise from the ordered-action threshold
representation. For example, to support $y=(1,0)$ as a Nash equilibrium,
player 1 must find entry optimal when player 2 stays out, so $\varepsilon_1\geq \underline{\epsilon}_1^\theta(x,\lambda).$ At the same time, player 2 must find staying out optimal when player 1
enters, so $\varepsilon_2\leq \overline{\epsilon}_2^\theta(x,\lambda).$
Thus the singleton-event region for $y=(1,0)$ is the rectangle
\[
\left[
\underline{\epsilon}_1^\theta(x,\lambda),+\infty
\right]
\times
\left[
-\infty,\overline{\epsilon}_2^\theta(x,\lambda)
\right],
\]
up to boundary points of probability zero. The same logic gives the
singleton-event regions for the other action profiles:
\[
\begin{aligned}
(0,0):\quad
&\varepsilon_1\leq \underline{\epsilon}_1^\theta(x,\lambda),
\quad
\varepsilon_2\leq \underline{\epsilon}_2^\theta(x,\lambda),\\
(1,1):\quad
&\varepsilon_1\geq \overline{\epsilon}_1^\theta(x,\lambda),
\quad
\varepsilon_2\geq \overline{\epsilon}_2^\theta(x,\lambda),\\
(0,1):\quad
&\varepsilon_1\leq \overline{\epsilon}_1^\theta(x,\lambda),
\quad
\varepsilon_2\geq \underline{\epsilon}_2^\theta(x,\lambda).
\end{aligned}
\]
Under logistic shocks, these rectangles yield the singleton-event
inequalities reported in Section \ref{section:unordered.actions}. Thus, with
binary actions, the unordered-logit and ordered-logit representations deliver
the same singleton-event probabilities, up to notation.

\subsection{Numerical Illustration: Multi-Store Entry Game}

I next consider a multi-store entry game in the spirit of
\citet{aradillas2022inference}. Let
$\mathcal Y_i=\{0,1,\ldots,L_i\}$ denote the set of possible store counts for
firm $i=1,\ldots,I$. I focus on the two-player case, $I=2$, and specify
payoffs as
\begin{equation}
\label{equation:chain.entry.game.payoff.function}
u_i(y_i,y_{-i},x,\lambda,\varepsilon_i)
=
y_i\left(
\beta_0+\beta_1x_i-\Delta_i y_{-i}-\eta y_i
+\sigma\lambda+\varepsilon_i
\right),
\end{equation}
where $\Delta_i>0$ and $\eta>0$. The parameter $\Delta_i$ captures the
competitive effect of the rival's stores on firm $i$'s payoff, while $\eta$
captures diminishing returns to firm $i$'s own stores.

This payoff can be written in the ordered-action form
\[
u_i(y_i,y_{-i},x,\lambda,\varepsilon_i)
=
V_i(y_i,y_{-i},x,\lambda)+y_i\varepsilon_i,
\]
where
\[
V_i(y_i,y_{-i},x,\lambda)
=
y_i\left(
\beta_0+\beta_1x_i-\Delta_i y_{-i}-\eta y_i+\sigma\lambda
\right).
\]
The deterministic adjacent payoff increment from increasing firm $i$'s action
from $y_i-1$ to $y_i$ is
\[
\beta_0+\beta_1x_i-\Delta_i y_{-i}
+\sigma\lambda
-
\eta(2y_i-1).
\]
This expression decreases strictly in $y_i$ whenever $\eta>0$. Hence
$V_i(\cdot,y_{-i},x,\lambda)$ is strictly concave on the ordered action grid.
The term $y_i\varepsilon_i$ implies that higher realizations of
$\varepsilon_i$ make higher actions more attractive. The restriction
$\Delta_i>0$ implies strategic substitutes by making the threshold higher
when the rival chooses more stores, but this comparative-static property is
not needed for the ordered-logit likelihood or the convexity result.

For $y_i=1,\ldots,L_i$, the threshold between actions $y_i-1$ and $y_i$ is
\begin{equation}
\label{equation:chain.entry.game.threshold}
e_i^\theta(y_i,y_{-i},x,\lambda)
=
\eta(2y_i-1)
+
\Delta_i y_{-i}
-
\beta_0
-
\beta_1x_i
-
\sigma\lambda,
\end{equation}
with boundary definitions
\[
e_i^\theta(0,y_{-i},x,\lambda)=-\infty,
\qquad
e_i^\theta(L_i+1,y_{-i},x,\lambda)=+\infty.
\]
Therefore firm $i$'s action $y_i$ is a best response to $y_{-i}$ if and only
if
\[
e_i^\theta(y_i,y_{-i},x,\lambda)
\leq
\varepsilon_i
\leq
e_i^\theta(y_i+1,y_{-i},x,\lambda).
\]
Under standard logistic shocks, the corresponding player-level
best-response probability is
\[
F\!\left(e_i^\theta(y_i+1,y_{-i},x,\lambda)\right)
-
F\!\left(e_i^\theta(y_i,y_{-i},x,\lambda)\right).
\]
Thus, for any candidate profile $y$, the conditional singleton-event
generalized likelihood is
\[
\mathcal L_{\theta}(y\mid x,\lambda)
=
\prod_{i=1}^2
\left[
F\!\left(e_i^\theta(y_i+1,y_{-i},x,\lambda)\right)
-
F\!\left(e_i^\theta(y_i,y_{-i},x,\lambda)\right)
\right].
\]

\begin{figure}[htbp!]
    \centering
    \includegraphics[width=0.55\textwidth]{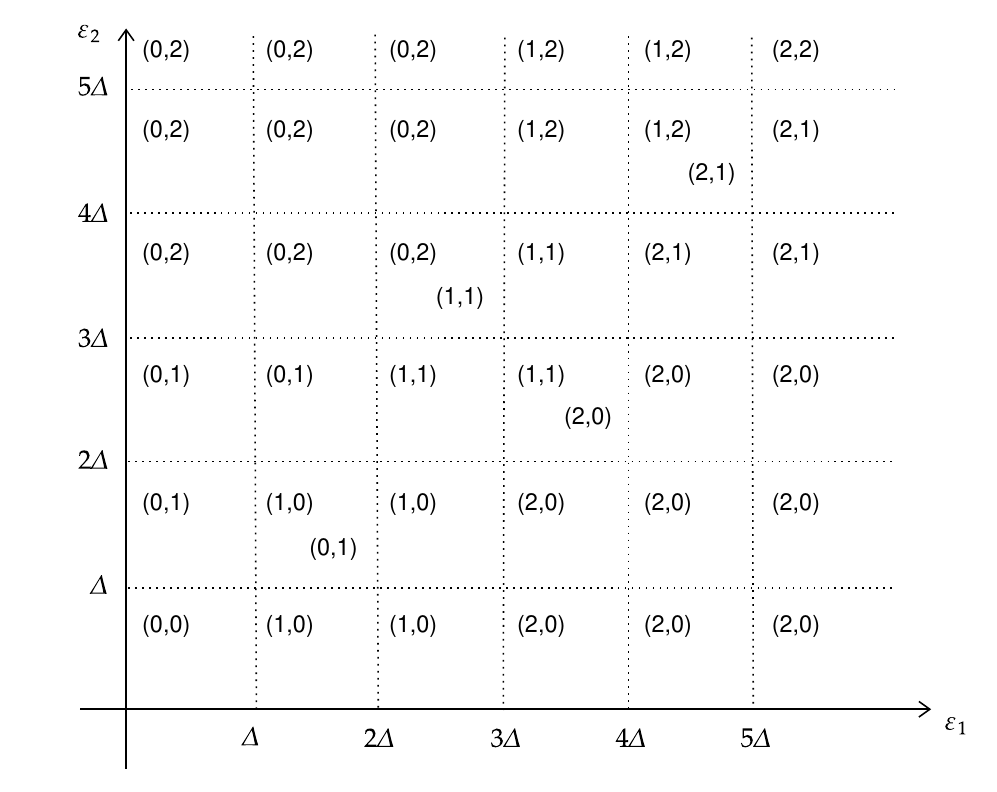}
    \caption{Structure of equilibria in a two-player multi-store entry game}
    \label{figure:ordered.actions.example.equilibria}
    \floatfoot{\emph{Notes:} This figure illustrates the structure of pure-strategy
Nash equilibria in a two-player, multi-store entry game with
$L_1=L_2=2$, $\beta_0=\beta_1=\sigma=0$, and
$\Delta_1=\Delta_2=\eta\equiv\Delta$. Each cell reports the set of
pure-strategy Nash equilibria for values of $(\varepsilon_1,\varepsilon_2)$
in that region, ignoring measure-zero boundary ties. The horizontal and
vertical bin labels partition the shock space using the threshold values
$\Delta,2\Delta,3\Delta,4\Delta$, and $5\Delta$. The leftmost column and
bottom row extend to the lower tails of the shock distribution, while the
rightmost column and top row extend to the upper tails.}
\end{figure}

Figure \ref{figure:ordered.actions.example.equilibria} illustrates the
structure of pure-strategy equilibria in a special case with
$L_1=L_2=2$, $\beta_0=\beta_1=\sigma=0$, and
$\Delta_1=\Delta_2=\eta\equiv\Delta$. In this case, the two finite
thresholds for player $i$, conditional on the rival's action $y_{-i}$, are
\[
e_i^\theta(1,y_{-i})=(1+y_{-i})\Delta,
\qquad
e_i^\theta(2,y_{-i})=(3+y_{-i})\Delta.
\]
Ignoring boundary ties, firm $i$'s best response is therefore
\[
BR_i(y_{-i};\varepsilon_i)
=
\begin{cases}
0,
& \varepsilon_i < (1+y_{-i})\Delta, \\[0.3em]
1,
& (1+y_{-i})\Delta < \varepsilon_i < (3+y_{-i})\Delta, \\[0.3em]
2,
& \varepsilon_i > (3+y_{-i})\Delta.
\end{cases}
\]
Because $y_{-i}\in\{0,1,2\}$, the relevant grid lines in the
$(\varepsilon_1,\varepsilon_2)$ plane occur at
\[
\Delta,\ 2\Delta,\ 3\Delta,\ 4\Delta,\ 5\Delta.
\]
Each cell in Figure \ref{figure:ordered.actions.example.equilibria} is
obtained by checking the mutual best-response conditions
\[
y_1\in BR_1(y_2;\varepsilon_1),
\qquad
y_2\in BR_2(y_1;\varepsilon_2).
\]
Some cells contain multiple pure-strategy equilibria because actions are
strategic substitutes.

The figure also illustrates why singleton-event probabilities are tractable.
For a fixed candidate profile $y$, the event that $y$ is a pure-strategy Nash
equilibrium is a rectangle in the shock space:
\[
\prod_{i=1}^2
\left[
e_i^\theta(y_i,y_{-i}),\,
e_i^\theta(y_i+1,y_{-i})
\right],
\]
again up to boundary points. For example, the singleton-event region for
$y=(1,0)$ is
\[
(\varepsilon_1,\varepsilon_2)
\in
[\Delta,3\Delta]\times(-\infty,2\Delta].
\]
Its conditional generalized likelihood in this simplified case is therefore
\[
\mathcal L_\theta((1,0)\mid x,\lambda)
=
\left[F(3\Delta)-F(\Delta)\right]F(2\Delta),
\]
where $x$ and $\lambda$ are suppressed because $\beta_1=\sigma=0$ in the
illustration.

I next simulate data from the same two-player multi-store entry game. I set
$L_1=L_2=2$ and impose symmetry in the competitive effects,
$\Delta_1=\Delta_2=\Delta$. The true parameter values are reported in Table
\ref{table:ordered.actions.simulation}. I set
$\sigma=0.5$ and assume that the econometrician knows this value to simplify
the simulation exercise. I draw each $\varepsilon_i$ independently from the
standard logistic distribution and draw $\lambda$ from the standard normal
distribution. I set each covariate $x_i$ to have binary support
$\{-1,1\}$. In the simulation, I compute conditional choice probabilities
for each of the four covariate profiles
$(x_1,x_2)\in\{-1,1\}^2$. Equivalently, the covariates can be interpreted as
independently drawn from a balanced two-point distribution,
$\Pr(x_i=-1)=\Pr(x_i=1)=1/2$, although the implementation conditions on each
covariate cell rather than drawing covariates market by market.

I generate 100,000 simulation draws for each covariate profile. For each
simulated market, I enumerate the set of pure-strategy Nash equilibria. When
multiple equilibria exist, the data-generating process selects uniformly from
the equilibrium set. This selection rule is used only to generate simulated
outcomes; it is not imposed when computing the singleton-event outer set.

Table \ref{table:ordered.actions.simulation} reports projection intervals for
the singleton-event outer set, along with the runtime for the singleton-set
projection step after the simulated conditional choice probabilities are
constructed. The intervals contain the true parameter values and are informative
in this simulation. The singleton-set projection computation takes 2.74 seconds.

\begin{table}[!htbp]
\small
\centering
\caption{Numerical simulation of a multi-store entry game}
\label{table:ordered.actions.simulation}
\begin{tabular}{lcc}
\toprule
Parameter & True value & Projection bounds \\
\midrule
$\beta_0$ & 0.5 & $[0.456,\,0.502]$ \\
$\beta_1$ & 1.0 & $[0.982,\,1.024]$ \\
$\Delta$  & 0.5 & $[0.462,\,0.501]$ \\
$\eta$    & 0.5 & $[0.487,\,0.506]$ \\
\midrule
Runtime (sec.) & & 2.74 \\
\bottomrule
\end{tabular}

\vspace{0.5em}
\begin{minipage}{0.75\textwidth}
\footnotesize
\emph{Notes:} The table reports projection intervals for the
singleton-event outer set in a two-player multi-store entry game.
The simulation uses 100,000 draws for each covariate profile.
The common-shock coefficient is fixed at $\sigma=0.5$ and treated
as known when computing the projection bounds. The singleton-event likelihood integrates over the common shock using
\(S=20\) deterministic normal-quantile quadrature points. Projection bounds are
computed using a log-inequality tolerance of \(0.003\), with the projection
level adjusted upward if the minimum numerical violation exceeds this tolerance.
Runtime is the singleton-set computation time after the simulated conditional
choice probabilities are constructed.
\end{minipage}
\end{table}

%% file: S.D.Computational.Details.tex
\section{Computational Details \label{section:computational.details}}

This appendix summarizes the numerical implementation used in the two empirical applications. Throughout, I solve fixed-\(\sigma\) problems and report projection intervals as unions over retained values of the common-shock scale. Payoff coefficients are reported in the normalized scale
\[
    \gamma
    =
    \theta/s(\sigma),
    \qquad
    s(\sigma)
    =
    \sqrt{\pi^2/3+\sigma^2},
\]
where \(\pi^2/3\) is the variance of a standard logistic shock. The model-implied generalized likelihoods integrate over a one-dimensional common market-level shock:
\[
    \mathcal L_{\theta}(A\mid x)
    =
    \int
    \mathcal L_{\theta}(A\mid x,\lambda)h(\lambda)d\lambda .
\]
I approximate this integral by deterministic quantile quadrature,
\[
    \mathcal L_{\theta}(A\mid x)
    \approx
    \frac{1}{S}
    \sum_{s=1}^{S}
    \mathcal L_{\theta}(A\mid x,\lambda^s),
    \qquad
    \lambda^s
    =
    H^{-1}
    \left(
        \frac{2s-1}{2S}
    \right),
\]
where \(H\) is the standard normal distribution function. I use \(S=20\) in the first empirical application and \(S=40\) in the burger-chain application.\footnote{The log-concavity results in the main text apply to the exact integrated likelihood. The finite quadrature approximation is the numerical implementation of that exact problem.}

\subsection{First Empirical Application}
\label{app:computation.first.application}

The first application is the two-player Walmart--Kmart entry game with action
space \(\mathcal Y=\{0,1\}^2\). I first group markets by the discretized
covariate support used in the main text. The continuous covariates
population, retail sales per capita, urban share, and distance to Bentonville
are discretized by median split and replaced by within-bin means; the South and
Midwest indicators are retained as discrete covariates. For each covariate cell
\(x\), let \(N_{y,x}\) be the number of markets with outcome \(y\), and let
\(n_x=\sum_{y\in\mathcal Y}N_{y,x}\).

For inference, I use Jeffreys-regularized within-cell frequencies,
\[
    \widehat p_x(y)
    =
    \frac{N_{y,x}+1/2}{n_x+|\mathcal Y|/2}.
\]
For an event \(A\subseteq\mathcal Y\), define
\[
    \widehat\phi(A\mid x)
    =
    \sum_{y\in A}\widehat p_x(y).
\]
The same regularized probabilities are used to simulate the max-statistic
critical value and to construct the log lower bounds entering the moment
inequalities.

For an event collection \(\mathcal J\), I simulate the covariance-aware
max statistic for the vector of log event probabilities. For each cell \(x\),
the multinomial Gaussian approximation has covariance
\[
    \operatorname{diag}(\widehat p_x)-\widehat p_x\widehat p_x' .
\]
Let \(c_{1-\alpha}\) be the simulated \(1-\alpha\) quantile, using
\(\alpha=0.01\) and 50,000 simulation draws. The resulting one-sided lower
bound for event-cell pair \((A,x)\) is
\[
    \ell(A,x)
    =
    \log \widehat\phi(A\mid x)
    -
    c_{1-\alpha}
    \sqrt{
        \frac{
            1-\widehat\phi(A\mid x)
        }{
            n_x\widehat\phi(A\mid x)
        }
    },
\]
with numerically degenerate denominators omitted. The implemented confidence
inequality is
\[
    \ell(A,x)
    \leq
    \log \mathcal L_{\theta}(A\mid x).
\]

For the singleton-event outer confidence set, \(\mathcal J\) contains the four
singleton action profiles in each covariate cell. For each fixed
\(\sigma\), I solve
\[
\begin{aligned}
    \min_{\gamma,t}\quad & t \\
    \text{s.t.}\quad
    &
    \ell(\{y\},x)
    \leq
    \log
    \mathcal L_{s(\sigma)\gamma}(\{y\}\mid x)
    +
    t,
    \qquad
    y\in\mathcal Y,\ x\in\mathcal X,\\
    & t\geq 0,\\
    & \gamma_{\Delta}\leq 0,
\end{aligned}
\]
where \(\gamma_{\Delta}\) is the competitive-effect coefficient. I scan
\[
    \sigma\in\{0,0.5,1.0,\ldots,10.0\}.
\]
A value of \(\sigma\) is retained if the minimized violation is no larger than
\(10^{-5}\), up to a \(10^{-8}\) numerical buffer. At each retained
\(\sigma\), I compute lower and upper projections for each component of
\(\gamma\) subject to the same inequalities and \(t\leq 10^{-5}\). The
reported singleton-event intervals are unions over retained fixed-\(\sigma\)
projection intervals.

I also compute a sampled approximation to the sharp/core confidence set. In the
two-player binary-action game, the core-determining event collection is
\[
    \mathcal A^*
    =
    \left\{
    \{(0,0)\},
    \{(0,1)\},
    \{(1,0)\},
    \{(1,1)\},
    \{(0,1),(1,0)\}
    \right\}.
\]
I recompute the max-statistic critical value for this larger event collection.
Because the sharp/core projection problem is not solved by convex optimization,
I use the singleton-event projection intervals to define sigma-specific search
boxes. These boxes are expanded slightly for the numerical search.

For each \(\sigma\) with a singleton-event search box, the sharp/core script
first minimizes the core violation criterion using a derivative-free global
search. If an accepted point is found, it then launches independent adaptive
random-walk chains initialized at that point. In the reported run, the
derivative-free search uses 60,000 steps and population size 300; the adaptive
stage targets \(10^6\) accepted points per \(\sigma\), split across six worker
processes. The reported sharp/core intervals are the coordinate-wise minima and
maxima of accepted points, unioned over \(\sigma\). They should therefore be
interpreted as inner numerical approximations to the sharp confidence-set
projections and may be narrower than the exact sharp-set projection intervals
if the sampler misses boundary points.

For the point-identified benchmarks, I estimate complete-information logit
models on the same discretized covariate support. The grouped cell-level
likelihood is likelihood-equivalent to expanding the discretized market-level
data, but is faster and exactly aligned with the partial-identification input
data. The benchmark coefficients are reported in the same normalized payoff
scale.

\subsection{Second Empirical Application}
\label{app:computation.burger}

The second application is the three-player ordered-entry game among
McDonald's, Burger King, and Wendy's. I begin from the 2019 cross section of
the Data Axle Historical Business Database for SIC code 58, which covers eating
and drinking places. I identify McDonald's, Burger King, and Wendy's outlets
using parent-company identifiers together with chain-name filters, and then
aggregate outlets to Census tracts. I merge these outlet counts with tract-level
covariates from the Food Access Research Atlas, Census tract coordinates, and
NaNDA eating-place counts. The final sample consists of urban Census tracts with
nonmissing covariates. Each chain's action is the number of outlets in the tract,
capped at \(2\), so \(2\) denotes two or more outlets.

Each chain's action is in \(\{0,1,2\}\), so the joint action space is
\[
    \mathcal Y=\{0,1,2\}^3 .
\]
The empirical CCP cells are formed from four discretized covariates: the number
of eating places in the tract and the three chain-specific leave-one-tract-out
county outlet densities, measured per 100,000 county residents. Each covariate
is discretized by a median split and replaced by its within-bin mean, yielding
\(16\) covariate cells. 

The burger-chain application uses the singleton-event outer confidence set. For
each \(y\in\mathcal Y\) and \(x\in\mathcal X\), let \(N_{y,x}\) be the raw
outcome count and let \(n_x=\sum_{y\in\mathcal Y}N_{y,x}\). I construct
simultaneous one-sided exact lower bounds for singleton probabilities using a
Bonferroni correction,
\[
    \alpha_{\mathrm{eff}}
    =
    \frac{\alpha}{|\mathcal Y||\mathcal X|}.
\]
For \(N_{y,x}>0\), the Clopper--Pearson lower bound is
\[
    \underline\phi(y\mid x)
    =
    F^{-1}_{\mathrm{Beta}
    \left(
        N_{y,x},
        n_x-N_{y,x}+1
    \right)}
    \left(
        \alpha_{\mathrm{eff}}
    \right),
\]
and for \(N_{y,x}=0\), I set \(\underline\phi(y\mid x)=0\). Singleton
inequalities with zero lower bound are nonbinding and are omitted from the log
formulation. The implemented restriction is
\[
    \log \underline\phi(y\mid x)
    \leq
    \log L_\theta(y\mid x),
    \qquad
    \underline\phi(y\mid x)>0,
\]
where \(L_\theta(y\mid x)\) is the model-implied probability that \(y\) is a
pure-strategy Nash equilibrium.

The ordered-action structure gives a closed-form expression for the conditional
best-response probabilities. Let
\[
    \delta_i(x)
    =
    \alpha_i+\beta_i E_x+\rho_i N_{i,x},
\]
where \(E_x\) is the number of eating places and \(N_{i,x}\) is chain \(i\)'s
own county-network variable. Directed competitive effects enter through
\[
    C_i(y_{-i})
    =
    \sum_{j\neq i}\Delta_{ij}y_j .
\]
Given \(y_{-i}\), \(x\), and \(\lambda\), the two finite thresholds for chain
\(i\)'s ordered action are
\[
    e_{i1}(y_{-i},x,\lambda)
    =
    \eta_i
    +
    C_i(y_{-i})
    -
    \delta_i(x)
    -
    \sigma\lambda,
\]
and
\[
    e_{i2}(y_{-i},x,\lambda)
    =
    3\eta_i
    +
    C_i(y_{-i})
    -
    \delta_i(x)
    -
    \sigma\lambda .
\]
Thus, conditional on \(\lambda\),
\[
    p_{i\theta}(y_i\mid y_{-i},x,\lambda)
    =
    \begin{cases}
    F_\varepsilon(e_{i1}), & y_i=0,\\
    F_\varepsilon(e_{i2})-F_\varepsilon(e_{i1}), & y_i=1,\\
    1-F_\varepsilon(e_{i2}), & y_i=2,
    \end{cases}
\]
where \(F_\varepsilon\) is the standard logistic distribution function. The
singleton generalized likelihood is
\[
    L_\theta(y\mid x)
    =
    \int
    \prod_{i=1}^{3}
    p_{i\theta}(y_i\mid y_{-i},x,\lambda)
    h(\lambda)d\lambda .
\]

For each fixed \(\sigma\), I solve
\[
\begin{aligned}
    \min_{\gamma,t}\quad & t \\
    \text{s.t.}\quad
    &
    \log \underline\phi(y\mid x)
    \leq
    \log
    L_{s(\sigma)\gamma}(y\mid x)
    +
    t,
    \qquad
    y\in\mathcal Y,\ x\in\mathcal X,\ 
    \underline\phi(y\mid x)>0,\\
    & t\geq 0,\\
    & \eta_i\geq 10^{-8},\qquad
      \Delta_{ij}\geq 0,\qquad i\neq j,\\
    & \gamma_k\in[-20,20]\qquad\text{for all }k .
\end{aligned}
\]
The box \([-20,20]\) is imposed for numerical stability and is not binding for
the reported projection intervals.

I scan over
\[
    \sigma\in\{0,0.5,1.0,\ldots,20.0\}.
\]
A value of \(\sigma\) is retained if the minimized violation is no larger than
\(10^{-5}\), up to a \(10^{-8}\) numerical buffer. For each retained value, I
compute projection bounds by minimizing and maximizing each component of
\(\gamma\) subject to the same inequalities and \(t\leq 10^{-5}\). Endpoint
problems are warm-started from the corresponding minimum-violation solution.
The reported intervals are unions over retained fixed-\(\sigma\) projection
intervals. The sigma scan and projection endpoints are parallelized across
Julia worker processes; reported runtime is total distributed computational
time.

\subsection{Computational Environment}

The data-preparation steps are implemented in R. The projection computations
are implemented in Julia using JuMP \citep{Lubin2023}; nonlinear programs are
solved with Ipopt \citep{wachter2006implementation}. All computations were run
on a 2024 24-inch iMac with an Apple M4 processor and 24 GB of RAM. Reported
runtimes are total computational times and should be interpreted as
implementation- and hardware-specific benchmarks. For distributed computations,
total computational time is the sum of runtimes across worker processes, rather
than elapsed wall-clock time.